%% file: main.tex
\documentclass[twocolumn,onecolappendix]{aastex63}

\usepackage{makecell}
\usepackage{xcolor}
\usepackage{amsmath,amsthm,amssymb,amsfonts}
\usepackage{graphics}
\usepackage{listings}
\usepackage{enumitem}
\usepackage{url}
\usepackage{appendix}
\usepackage[super]{nth}
\usepackage{mathtools}
\usepackage{hyperref}

\definecolor{dgreen}{RGB}{26,148,49}
\definecolor{xlinkcolor}{cmyk}{1,1,0,0}
\hypersetup{linkcolor=xlinkcolor,citecolor=xlinkcolor,urlcolor=xlinkcolor}

\newcommand{\N}{\mathbb{N}}
\newcommand{\Z}{\mathbb{Z}}

\begin{document}

\correspondingauthor{Richard M. Feder}
\author[0000-0001-5382-6138]{Richard M. Feder}
\email{rmfeder@berkeley.edu}
\affiliation{Berkeley Center for Cosmological Physics, University of California, Berkeley, CA 94720,
USA}
\affiliation{Lawrence Berkeley National Laboratory, Berkeley, California 94720, USA}

\author[0009-0007-4952-1674]{Liam Parker}
\affiliation{Berkeley Center for Cosmological Physics, University of California, Berkeley, CA 94720,
USA}
\affiliation{Lawrence Berkeley National Laboratory, Berkeley, California 94720, USA}

\author[0000-0001-5382-6138]{Uroš Seljak}
\affiliation{Berkeley Center for Cosmological Physics, University of California, Berkeley, CA 94720,
USA}
\affiliation{Lawrence Berkeley National Laboratory, Berkeley, California 94720, USA}

\title{A Probabilistic Autoencoder for Galaxy SED Reconstruction and Redshift Estimation: Application to Mock SPHEREx Spectrophotometry}
\begin{abstract}
We present a probabilistic autoencoder (PAE) framework for galaxy spectral energy distribution (SED) modeling and redshift estimation, applied to synthetic SPHEREx 102-band spectrophotometry. Our PAE learns a compact latent representation of rest-frame galaxy SEDs transformed to a simple Gaussian base density using a normalizing flow, combined with an explicit forward model enabling joint Bayesian inference over intrinsic SED parameters and redshift with well-defined priors. In controlled tests on simulated SPHEREx spectra, our PAE improves on template fitting (TF) in source recovery, outlier rate, and posterior calibration, with trade-offs in redshift performance that depend on the assumed priors. A simple cut on the ratio of PAE and TF uncertainties identifies sources that are overwhelmingly TF outliers, which can be used to clean existing TF samples while retaining the vast majority of well-recovered sources. By directly profiling over PAE latent variables, we show these cases correspond to shallow likelihood surfaces where the PAE's continuous SED manifold produces broader likelihoods that more faithfully reflect the lack of constraining power in the data, whereas the TF discrete model grid yields artificially confident but incorrect redshift estimates. Lastly, we present an alternative, simulation-based inference approach using a Transformer encoder and conditional normalizing flow, which provides similar redshift performance to the PAE but with $\sim200\times$ faster inference throughput. Our implementation, \texttt{PAESpec}, is publicly available and provides a foundation for principled redshift estimation in modern photometric surveys. 
\end{abstract}

\keywords{Cosmology; Large-scale structure of the universe; Galaxy evolution; Catalogs; Surveys}

\input{sections/introduction}
\input{sections/pae_model}
\input{sections/implementation}

\input{sections/sampling}
\input{sections/pae_tests}
\input{sections/redshift_inference}
\input{sections/sbi}
\input{sections/conclusion}
\input{sections/acknowledgements}

\clearpage
\appendix
\input{sections/convergence_app}

\bibliography{references}{}
\bibliographystyle{aasjournal}

\end{document}

%% file: sections/introduction.tex
\section{Introduction}
\label{sec:introduction}
Modern galaxy surveys are delivering unprecedented volumes of multi-wavelength photometry and low-resolution spectroscopy, enabling novel studies of galaxy evolution and large-scale structure (LSS). SPHEREx, NASA's most recent MIDEX mission, began survey operations in May 2025 and is conducting an all-sky near-infrared spectral survey spanning $0.75-5$ $\mu$m, with several hundred million spectra with spectral resolving power $R\equiv \lambda/\Delta \lambda =35-125$ \citep{bock25}. This spectrophotometric dataset occupies an intermediate regime between broad-band photometry and spectroscopy -- while the majority of SPHEREx galaxies will not have secure emission-line based redshifts (in contrast to existing spectroscopic surveys), the dense spectral sampling contains rich color and spectral energy distribution (SED) shape information that broad-band photometric redshifts only partially access. The immense science opportunities in cosmology and astrophysics afforded by SPHEREx and other multi-wavelength datasets, e.g., from \emph{Gaia} \citep{gaia}, \emph{Rubin} \citep{rubin_ivezic}, \emph{Euclid} \citep{euclid}, etc., place a premium on methods that can effectively model SEDs, produce well-calibrated photometric redshift (``photo-z") uncertainties and scale well to large samples \citep{rail}.

All photo-z methods balance trade-offs between accuracy, flexibility, robustness, and interpretability. Template fitting (denoted TF throughout) approaches, which fit a library of SEDs (or combinations of SEDs) to observed photometry using a well-defined forward model \citep{lephare, topz, eazy}, have become a workhorse for modern surveys given their computational efficiency and simplicity. However, TF methods that use purely model-based and/or mis-calibrated template sets may struggle to reproduce the full variability of galaxy SEDs captured by increasingly rich datasets. Furthermore, the profile likelihood approach used by many TF codes for $p(z)$ estimation (e.g., profiling over template index, dust law/strength, amplitude) can produce biased or overconfident redshift estimates when nuisance parameters are correlated with redshift or poorly constrained, which becomes more common in low signal-to-noise ratio (SNR) measurements.

Alternatively, data-driven photo-z methods using decision trees/random forests \citep{tpz, izbiki_lee_17}, nearest-neighbor approaches \citep{graham_rubin, dnf}, self-organizing maps \citep{speagle_som}, dictionary learning \citep{bryan23, frontera23}, flow-based density estimation \citep{pzflow}, and simulation-based inference \citep[SBI;][]{Cranmer2020, Hahn2022} have found application across a variety of datasets. These methods learn flexible mappings from observables to physical parameters in a manner that is often informed by high-quality training data. While powerful, they often lack a robust probabilistic framework, with implicit priors that depend strongly on the variance and representativeness of the training set. Hybrid approaches using Gaussian processes \citep{leistedt_hogg} have made progress in addressing gaps between TF and data-driven methods, though challenges remain. We refer the reader to \cite{salvato18} for a more detailed review of photo-z methods, and to \cite{rail_test} for a recent comparison of photo-z methods using early \emph{Rubin} imaging data.

In this work, we use the probabilistic autoencoder (PAE), first introduced in \cite{bohm_pae}, as the basis for a framework to model galaxy spectra and perform robust redshift estimation. Unlike variational autoencoders, which assume a Gaussian prior over the encoded latent distribution, our PAE decouples the density prior from the autoencoder training, allowing the model to first prioritize data reconstruction. After this, the latent variable distribution is transformed to a simpler base distribution in a second training stage using normalizing flows. PAEs have been successfully demonstrated as a competitive framework for modeling multi-band supernovae time series in \cite{stein_pae}. The application to photo-z estimation builds on the fact that galaxy spectra can be described to good approximation by a small number of latent parameters \citep{portillo_sdss}. Similar to the \texttt{Spender} model \citep{spender1}, we use the PAE to define a rest-frame SED model, after which we rely on physically-informed forward modeling to reconstruct the data \citep{lanusse_galaxy_gen}. Our method can be viewed as a generalization of template fitting, in which the discrete set of templates is replaced with a learned continuous SED manifold and explicitly controllable priors. Through controlled comparisons, we seek to understand differences between TF and PAE redshift estimates, along with their complementarity.

The paper is structured as follows. We first introduce the PAE model as a forward modeling tool for Bayesian inference in \S \ref{sec:pae}. After describing the synthetic dataset of galaxy spectra we use to generate SPHEREx spectrophotometry in \S \ref{sec:dataset}, we describe our implementation \texttt{PAESpec} and describe its computational performance in \S \ref{sec:implementation}. We evaluate the performance of our PAE on SED reconstruction in \S \ref{sec:results} and redshift estimation in \S \ref{sec:inference}, using TF as a baseline for close comparison. In \S \ref{sec:sbi} we compare our results with an alternative SBI approach, which offers amortized posteriors conditioned on noisy, observed photometry. We conclude in \S \ref{sec:conclusion} and discuss potential extensions of our framework for cross-survey photo-z estimation and hierarchical inference.


%% file: sections/pae_model.tex
\section{Probabilistic Autoencoder}
\label{sec:pae}

\begin{figure*}
    \centering
\includegraphics[width=0.49\linewidth]{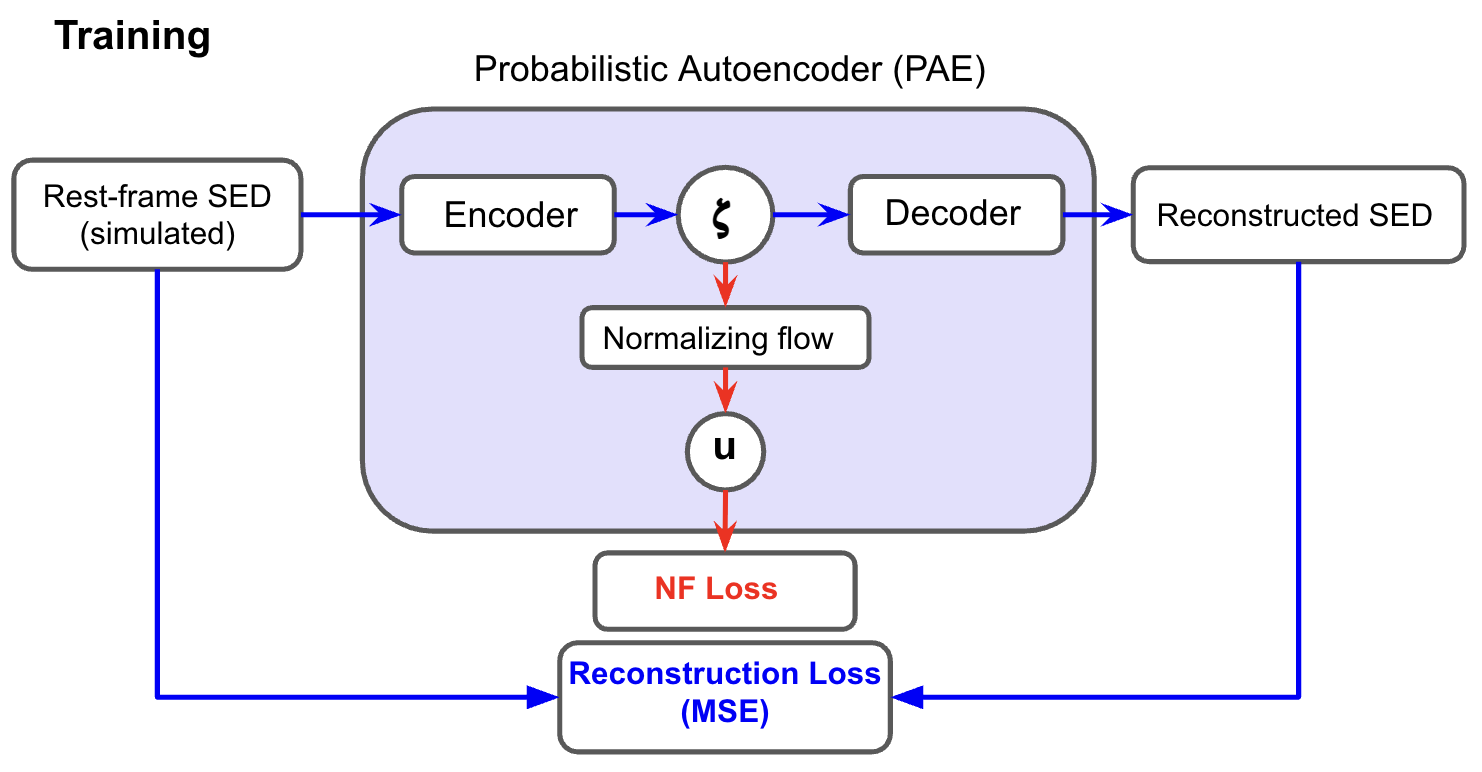}
\includegraphics[width=0.46\linewidth]{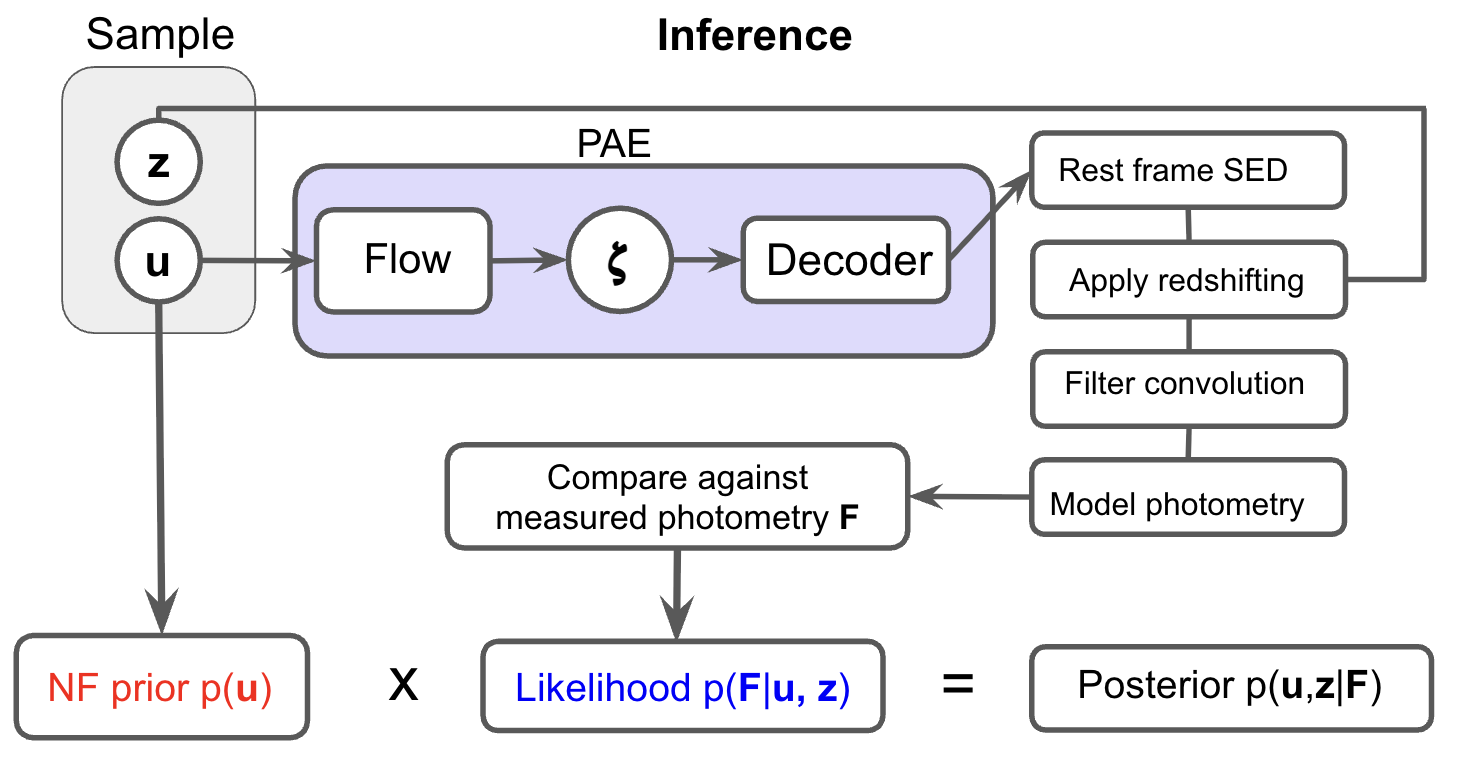}
    \caption{Schematic describing the training (left) and inference (right) stages of our probabilistic autoencoder model. The variables $\pmb{\zeta}$, $\mathbf{u}$ and $z$ denote the autoencoder latent vector, normalizing flow basis vector, and redshift, respectively. The blue and red arrows specify separate forward passes used when training the autoencoder and the normalizing flow. We note that additional priors beyond the NF prior (e.g., on redshift) can be folded into the PAE inference.}
    \label{fig:pae_schematic}
\end{figure*}

We train a probabilistic autoencoder (PAE) on synthetic, rest-frame galaxy spectra to learn a low-dimensional latent representation of spectral energy distributions (SEDs), independent of observational effects. We then combine this rest-frame SED model with explicit redshifting and filter convolution operations to predict observed photometry. While models such as \texttt{Spender} and its extension to quasars, \texttt{SpenderQ} \citep{hahn_spenderq}, have leveraged physically informed latent space models, they are focused on higher-resolution spectroscopy where redshifts are already known to high precision. Our approach is designed to enable joint inference of the galaxy's redshift and intrinsic SED. We outline the structure of the PAE schematically in Fig. \ref{fig:pae_schematic}.

\subsection{Notation}
For each source, we will have a collection of photometric measurements $\lbrace \lambda_i, f_i, \sigma_{i}\rbrace_{i=0}^{N_{\lambda}}$ where $N_{\lambda}$ is the total number of measurements per source at wavelengths $\lbrace \lambda_i \rbrace$. We define the following quantities, which are used throughout the text:
\begin{itemize}
    \item $\textbf{f}_{N_{\lambda}}$: Observed flux densities
    \item $\pmb{\sigma}_{N_{\lambda}}$: Observational uncertainties
    \item $\hat{\textbf{f}}_{N_{\lambda}}$: Reconstructed flux densities. 
    \item $\mathbf{s}$: High-resolution (normalized) spectrum.
    \item $z$: Galaxy redshift.
    \item $\mathbf{\zeta}_{n_{\rm latent}}$: Autoencoder latent parameters.
    \item $\textbf{u}_{n_{\rm latent}}$: Latent space parameter vector after transformation by the normalizing flow.
\end{itemize}

\subsection{Latent SED Model}
An autoencoder compresses high-dimensional data samples into a lower-dimensional representation $\vec{\zeta}$ using an encoder network, $e(\cdot)$, and reconstructs them with a decoder $d(\cdot)$,
\begin{equation}
    \mathbf{s}_{\rm rest} = d(e(\mathbf{s})).
\end{equation}
Training the autoencoder through this ``information bottleneck" (i.e., where $n_{\rm latent} \ll N_{\lambda}$) encourages the model to learn a compact representation that captures the salient features of the data, and can be thought of as a form of non-linear PCA \citep{portillo_sdss}.

We first train the autoencoder to reconstruct noiseless, rest-frame spectra $\textbf{s}$. Unlike the more commonly-used variational autoencoder, which regularizes the reconstruction loss with a Gaussian prior over latent parameters $\pmb{\zeta}$, we impose no such prior at this stage, letting the autoencoder determine a representation that best captures the ``explained variance" of the data \citep{explained_var}. This also means that the multivariate distribution $p(\mathbf{\zeta})$ may be complex.

We then train a normalizing flow (NF) to map the empirical distribution of latent vectors $p(\pmb{\zeta})$ to a simpler base distribution $p(\textbf{u})$, typically chosen to be a standard Gaussian. The NF model is bijective by construction, enabling exact inversion between $\pmb{\zeta}$ and $\textbf{u}$. This transforms the challenge of sampling a complex, potentially multi-modal latent distribution into the simpler task of sampling from $p(\textbf{u})$. We note that while the flow is trained to match a standard Gaussian, residual structure in the transformed space may persist due to the capacity of the flow model relative to complexity of the learned SED manifold.

\subsection{Forward Modeling to Observed Photometry}
Given a rest-frame spectrum $\mathbf{s}_\mathrm{rest}$, the observed spectrum at redshift $z$ is obtained through a wavelength rescaling,
\begin{equation}
    \mathbf{s}_\mathrm{obs}(\lambda) = \mathbf{s}_\mathrm{rest}\left(\frac{\lambda}{1+z}\right).
\end{equation}
For band $i$ with filter $T_i(\lambda)$, the convolved model flux $\tilde{f}_i$ is given by
\begin{equation}
    \tilde{f}_i = \frac{\int T_i(\lambda) \mathbf{s}_{\rm obs}(\lambda) \lambda \, d\lambda}{\int T_i(\lambda) \lambda \, d\lambda}.
\end{equation}
This is similar to the \texttt{Spender} model \citep{spender1}; however, while \texttt{Spender} passes observed-frame (i.e. redshifted) spectra through the encoder for direct point estimates, we do not utilize the encoder beyond the initial PAE training, instead using the decoder + forward model for inference.

Rather than apply a fixed normalization to the observed spectra, we include an overall scale factor $A$ in our forward model. For Gaussian errors and a flat prior on $A$, the scale factor can be marginalized analytically, with best-fit $A_*$ given by
\begin{equation}
    A_* = \frac{\sum_i f_i \tilde{f}_i/\sigma_i^2}{\sum_i \tilde{f}_i^2/\sigma_i^2}.
\end{equation}
We found that the marginalization step is important for obtaining unbiased reconstructions of the high-SNR sources in our sample, compared to fixing both the data and model SED normalizations.

\subsection{Posterior analysis}
To perform inference, we sample from the PAE posterior distribution over SED latent parameters and redshift, conditioned on the observed photometry. A benefit of our framework is the ability to cleanly define the prior structure and test the impact of different priors on our resulting photo-z estimates, which we explore in \S \ref{sec:prior_impact}.

Our likelihood, $\ln P(\mathbf{f}|\mathbf{u}, z)$, assumes independent Gaussian errors and is evaluated using the reconstructed fluxes $\hat{f}_i(\mathbf{u}, z)$ from the forward model:
\begin{equation}
\ln P(\mathbf{f}|\mathbf{u}, z)= -\frac{1}{2} \sum_{i=1}^{N} \left( \frac{f_i - \hat{f}_i(\mathbf{u}, z)}{\sigma_i} \right)^2 + \mathrm{const.}
\end{equation}
The trained normalizing flow defines a prior over the training data encoded in latent variables $\textbf{u}$, which enables efficient probabilistic reconstruction from the reduced-dimension representation. We assume a model that factorizes latent parameters and the physical redshift, i.e., $d(\pmb{\zeta}, z) \sim p(\mathbf{u}, z) = p(\mathbf{u})p(z)$. This neglects the correlations between redshift and galaxy SEDs, which persist in the latent space even when de-coupling the physical redshift parameter \citep{spender2}\footnote{A more constrained approach is to train the flow on paired $\lbrace \zeta, z \rbrace$, such that $P(\zeta, z) \sim P(\mathbf{u})$, where $P(\mathbf{u})$ follows a Gaussian prior. In this approach the prior $P(z)$ is learned automatically as part of the flow, however it is potentially too constraining if the training set is too small or unrepresentative.}. The joint posterior is given by:
\begin{equation}
\ln p(\mathbf{u}, z|\mathbf{f}) = \ln p(\mathbf{f}|\mathbf{u}, z) + \ln p(\mathbf{u}) + \ln p(z).
\end{equation}
For faint, low-SNR sources, the prior dominates the posterior, while for higher-SNR objects the inference is driven by the likelihood. We explore the impact of assumed priors in \S \ref{sec:prior_impact}.

%% file: sections/implementation.tex
\section{Mock Data}
\label{sec:dataset}
We utilize galaxy simulations that are presented in \cite{feder23} and publicly available on Zenodo\footnote{\url{https://zenodo.org/records/11406518}}. The simulations are based on template fits to thirty-band photometry for a sample of 166,014 sources with $18 < i < 25$ from the COSMOS-2020 catalog \citep{weaver22}. 

To train the PAE, we select all sources with $z_{\rm AB}<23$, corresponding to 31492 sources. We then use this as the basis for an augmented sample, in which we generate multiple copies of each source with perturbations in input redshift and extinction $E(B-V)$. Using this procedure we generate a sample of 200,000 spectra for training. We place the synthetic high-resolution SEDs onto a grid of 500 linearly spaced wavelength elements spanning $0.1 < \lambda < 5.0$ $\mu$m. This ensures adequate spectral coverage and resolution over the SPHEREx bandpass for redshifts $z < 4$.

For testing PAE posterior estimation, we use synthetic 102-band SPHEREx photometry from SEDs generated in a similar fashion as our training set. We add Gaussian noise consistent with the maximum expected value (MEV) point source sensitivity used in \cite{feder23}. Our assumed sensitivity is $\sim 0.1-0.2$ mag shallower than the achieved in-flight sensitivity reported in \cite{bock25}. We restrict our test sample to galaxies with $z_{\rm AB}<22.5$. This is a brighter cut than the multi-band selection used in \cite{feder23} (which incorporated optical/IR broad-band photometry into redshift estimates), and also differs from the fiducial $z$-band and $z-W1$ color selection adopted in \cite{huai25}.

\begin{figure}
    \centering
    \includegraphics[width=\linewidth]{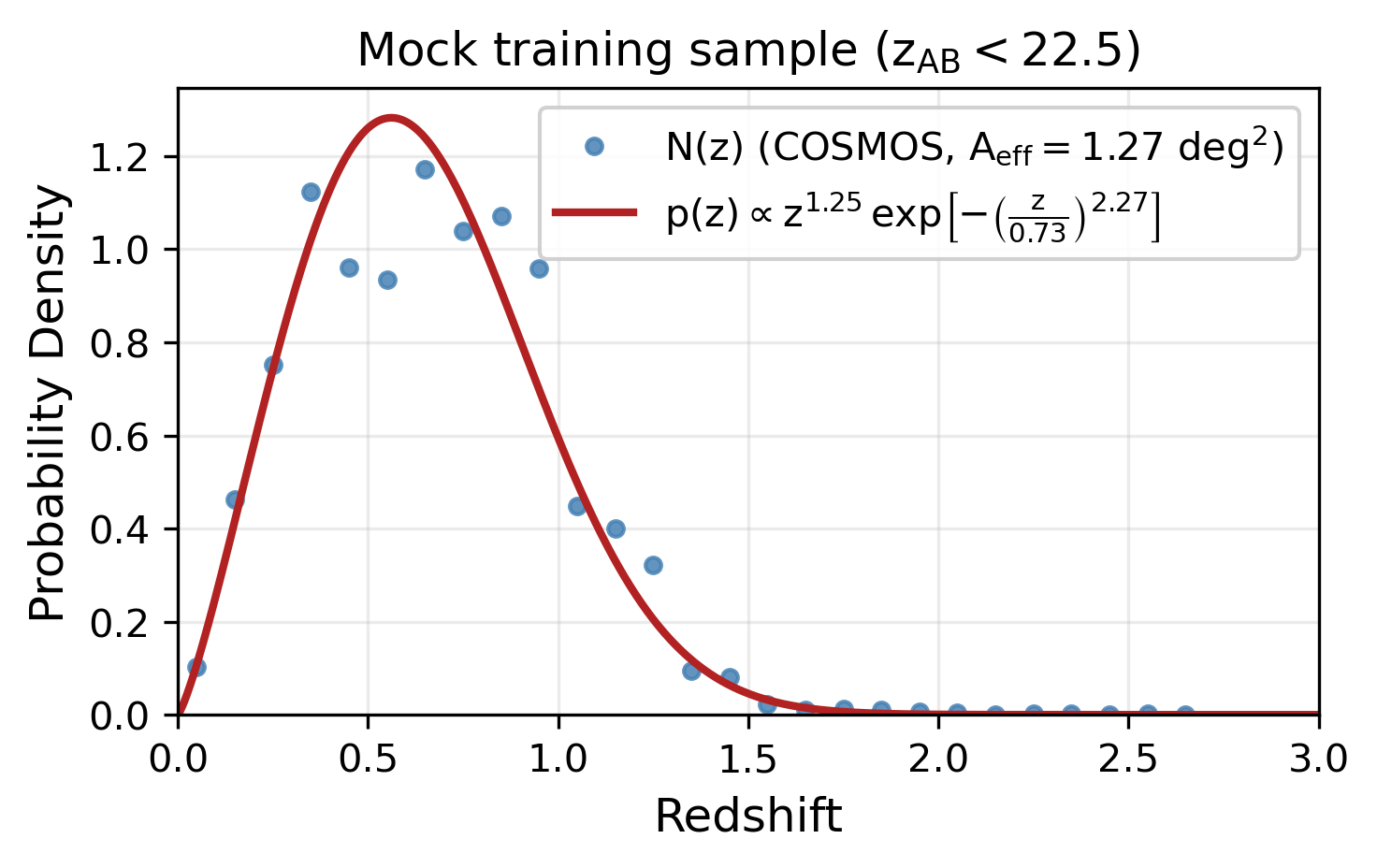}
    \caption{Number counts in $dz=0.1$ bins from our mock COSMOS sample (blue), alongside our best-fit parametric model for $p(z)$. Note that, due to the relatively small COSMOS-2020 footprint, sample variance uncertainties on the measured $N(z)$ are significant, i.e., our fitted prior is fairly approximate.}
    \label{fig:nzprior}
\end{figure}

In Figure \ref{fig:nzprior} we show the redshift distribution of the mock dataset, along with a smooth dN/dz model fit to the data. We assume the following parameterization for $N(z)$:
\begin{equation}
    p(z) \propto z^{\alpha}\exp\left[-\left(\frac{z}{\beta}\right)^{\gamma}\right],
\end{equation}
in which $\lbrace \alpha, \beta, \gamma\rbrace$ are free parameters fit to the data \citep{benitez_bpz}. We use this functional form to define our fiducial redshift prior, which we refer to as $p_{\rm BPZ}(z)$. This is chosen as a flexible prior for the redshift distribution with a unimodal distribution with an exponential tail towards higher $z$.

\section{Implementation}
\label{sec:implementation}

\subsection{Model and training}

\subsubsection{Autoencoder}
We employ a one-dimensional convolutional neural network (CNN) for both the encoder and decoder networks. The encoder compresses each rest-frame spectrum into a low-dimensional latent vector, while the decoder uses transpose convolution layers to reconstruct the input spectrum. Our encoder architecture consists of four convolutional layers with $[12, 32, 128, 256]$ filters and kernel sizes of $[5, 5, 5, 5]$, followed by a fully-connected network with $[256, 64, 16]$ hidden nodes. The decoder has a near identical but symmetric architecture. In total the autoencoder (AE) contains $\mathcal{O}$(200k) parameters. After experimenting with different data normalization schemes, we chose to mean-normalize the SEDs before inputting them into the encoder network. 
For training, we use the Adam optimizer \citep{adam} with an initial learning rate of $\eta=2\times 10^{-4}$ and batch size of 128. To help facilitate convergence, we use an exponential learning rate scheduler that decreases the learning rate by 10\% every five epochs. Our training further employs early stopping with a patience of 10 epochs and precision threshold of $10^{-3}$. We train the autoencoder over 100 epochs on the synthetic galaxy spectra described in \S \ref{sec:dataset}. We use an 80/20 training/validation split to monitor and prevent overfitting. 

To evaluate the impact of latent space dimensionality on our autoencoder reconstructions, we train four models with $n_{\rm latent} \in \lbrace 3, 5, 8, 10\rbrace$ and compare the reconstruction performance for SEDs in the validation dataset. We use the mean-squared-error (MSE) loss between the reconstructed spectrum and the input, i.e., 
\begin{equation}
    \mathcal{L}_{MSE} = \frac{1}{N}\sum_{i=1}^{N}(\hat{s}_i - s_i)^2,
\end{equation}
where $\hat{s}_i$ is the reconstructed spectrum for source $i$ and $s_i$ the input spectrum. As the SEDs are mean-normalized prior to encoding, the MSE is equivalent to the mean squared fractional deviation relative to the mean SED flux, while $\sqrt{\mathcal{L}_{\rm MSE}}$ gives the typical per-wavelength fractional reconstruction error.

\begin{figure*}[t]
    \centering

    \includegraphics[width=\linewidth]{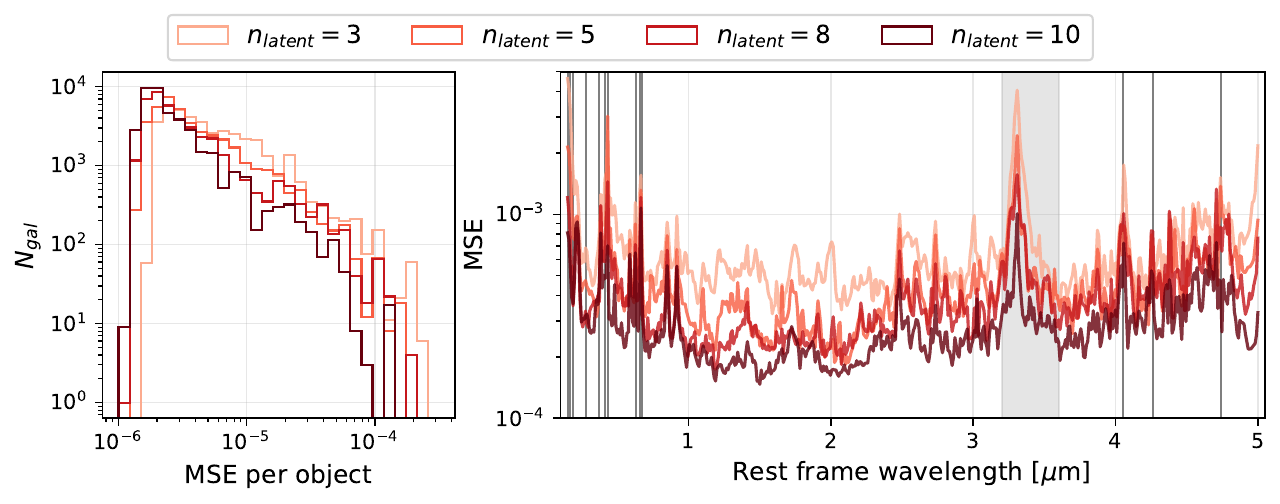}
    
    \caption{Rest-frame SED reconstruction performance of our trained autoencoder. The left panel shows the distribution of per-galaxy mean-squared errors (MSE) from our validation set, while the right panel shows the average MSE as a function of wavelength. The vertical lines correspond to the position of known emission lines, while the shaded band covers both the 3.3 $\mu$m and 3.4 $\mu$m aliphatic shoulder from polycyclic aromatic hydrocarbons (PAHs).}
    \label{fig:mse_recon_alltrain}
\end{figure*}

In Figure \ref{fig:mse_recon_alltrain} we show the results of this exercise, plotting both the per-object MSE and the MSE averaged over all objects as a function of rest-frame wavelength. The majority of our SEDs have high-fidelity reconstructions with fractional errors at the $\sim 1-3\%$ level, rising near wavelengths with strong non-linear features such as emission lines and the 3.3 $\mu$m polycyclic aromatic hydrocarbon (PAH) complex. As expected, the MSE decreases as we increase the latent dimension. In Appendix \ref{sec:vary_nlatent}, we further compare redshift recovery metrics, finding that performance continues to improve up to $n_{\rm latent}=10$. We therefore adopt this as our fiducial choice in \S \ref{sec:inference} when we evaluate redshift recovery.

\subsubsection{Normalizing flow}
A normalizing flow (NF) is a model defined by a series of learned bijective transformations,
\begin{equation}
    f = f_1 \cdot f_2 ... \cdot f_n
\end{equation}
that translate a base distribution $\pi_{\pmb{u}}$ to an arbitrary target density $\pi_{\pmb{\zeta}}$ \citep{normalizing_flows}. In our case, the target density is defined by the distribution of AE latent parameters $\pmb{\zeta}$. 

\begin{equation}
    \pi(\pmb{\zeta}) = \pi_u(f^{-1}(\pmb{\zeta}) \left|\det\left(\frac{\partial f^{-1}}{\delta \pmb{\zeta}}\right)\right|.
\end{equation}
The determinant describes the Jacobian of the inverse transformation from $\pmb{\zeta}$ to $\pmb{u}$. We seek a model where $p(\pmb{\zeta})$ is Gaussian distributed, which can be achieved by minimizing the Kullback-Leibler divergence between $p(\pmb{\zeta})$ and $\pi_{\pmb{\zeta}}$ or, equivalently, by maximizing the data likelihood $\log p(\pmb{\zeta})$. 

We use a multiple-layer inverse autoregressive flow \citep[IAF;][]{iaf} to transform a simple Gaussian latent prior $\pmb{u} \sim \mathcal{N}(0, I)$ into the more complex posterior over AE latent variables $\pmb{\zeta}$ through flow $f$. Each IAF block consists of a transformation parameterized by a masked autoregressive density estimator (MADE) neural network, which predicts the parameters of a flexible, monotonic rational quadratic spline \citep{duncan2019_nsf}. In place of the standard affine transformation, each latent variable $\zeta_i$ is transformed using a spline whose shape is conditioned on the preceding latent variables $\zeta_{<i}$. This transformation remains invertible and allows for fast sampling via the inverse mapping, enabling efficient PAE inference.

We apply an initial Z-score normalization to the input/output, motivated by the observation that, absent an explicit prior on $p(\pmb{\zeta})$ in the AE training, the learned latent distribution concentrates to a region with dispersion $\ll 1$. Before and after each IAF layer, we apply activation normalization (ActNorm) transformations with trainable per-feature scale and bias parameters that improve the training stability. We use randomized permutations of the features at each step to encourage dependencies across all latent dimensions.  

We build and train our NFs using the \texttt{FlowJAX} package \citep{ward2023flowjax}, which provides automatic scheduling and early stopping based on training and validation set losses. 

\begin{figure*}[t]
    \centering
    \includegraphics[width=0.48\linewidth]{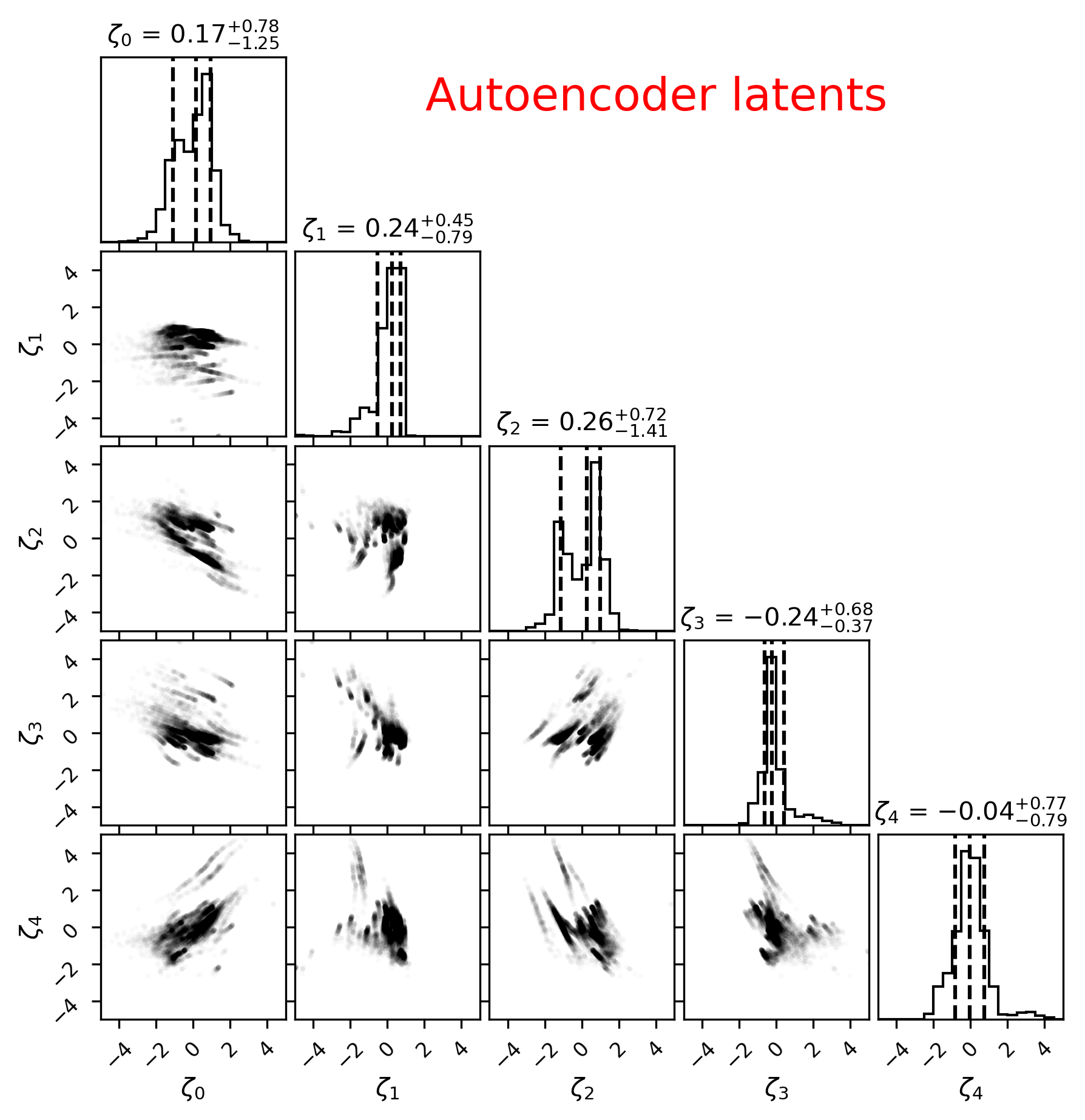}
    \includegraphics[width=0.48\linewidth]{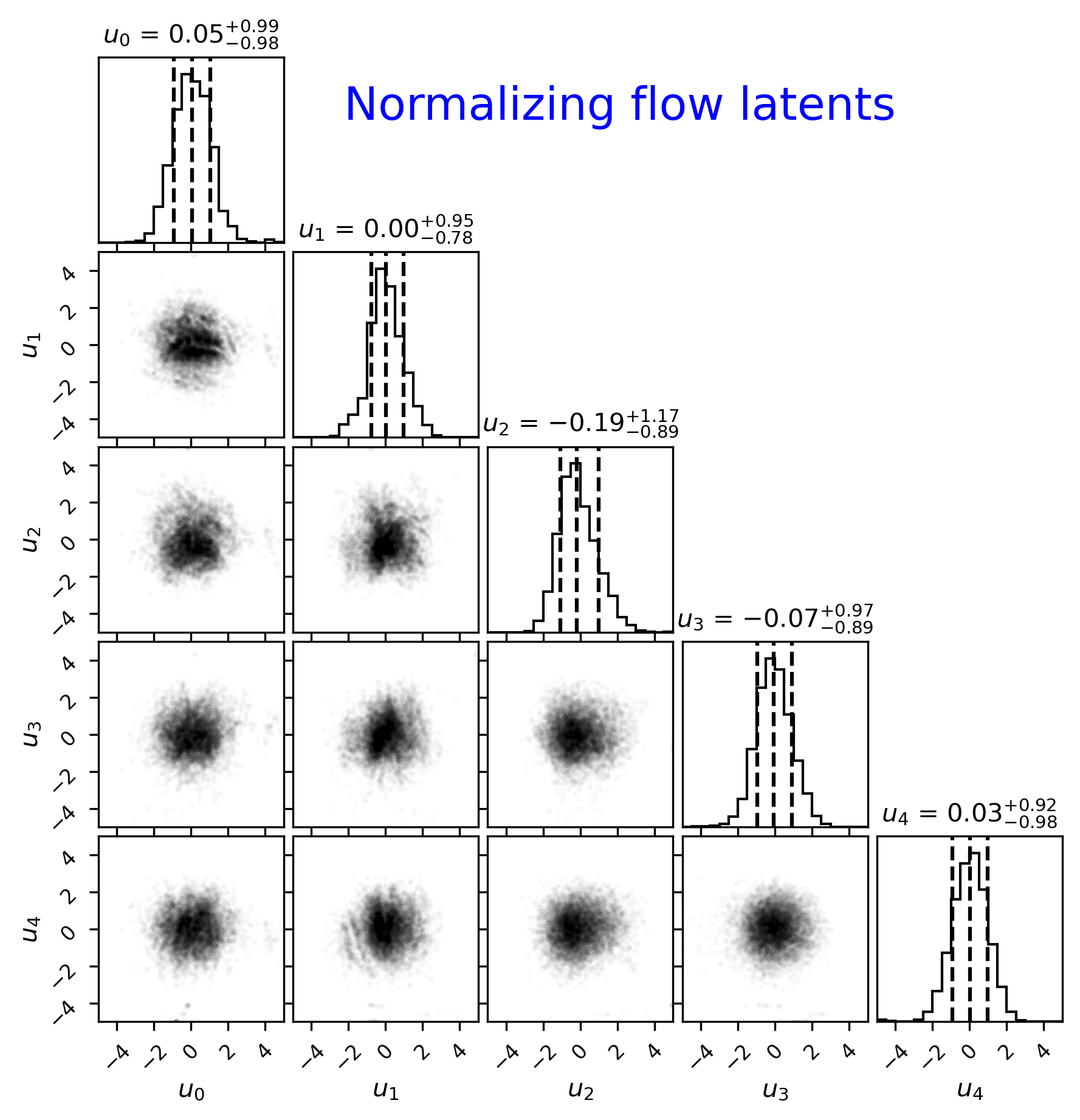}
    \caption{An example comparison of the autoencoder latent variable distribution $p(\pmb{\zeta})$ (left) and normalizing flow latent distribution $p(\mathbf{u})$, for the case $n_{\rm latent}=5$. The normalizing flow transforms $p(\pmb{\zeta})$, which is highly non-Gaussian and multimodal, into a simpler base distribution, though some residual structure remains.}
    \label{fig:latents_dist}
\end{figure*}

In Figure \ref{fig:latents_dist} we compare the distribution of autoencoder latent variables as well as the transformed normalizing flow distribution. One clearly sees that, without a variational prior, the autoencoder learns a highly non-Gaussian distribution that features multiple ``tracks" stemming from the underlying template set used to generate the SEDs. While some outliers remain in the normalizing flow latents, the base density is much closer to Gaussian-distributed (with all first and second moments matching expectations within 20\%), making it easier to explore the model space during sampling. 

\begin{figure}
    \centering
    \includegraphics[width=\linewidth]{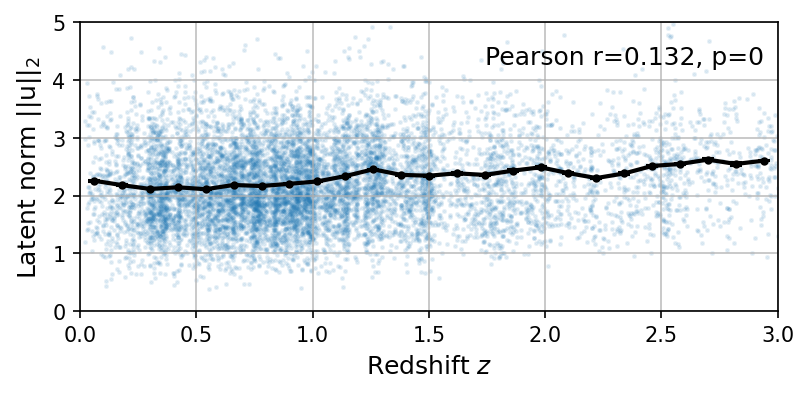}
    \caption{Normalizing flow latent vector norms for 10000 galaxies in our validation set as a function of redshift. The mean trend (black) shows a mild dependence of $||u||_2$ with redshift between $z=0$ and $z=3$, with Pearson correlation coefficient $r=0.132$.}
    \label{fig:latent_u2_vs_z}
\end{figure}

To test the variation of the PAE latent parameter distribution with redshift, in Fig. \ref{fig:latent_u2_vs_z} we show the distribution of NF latent vector norms as a function of redshift for 10000 spectra in the validation set spanning $0<z<3$. We observe a weak correlation between $||u||_2$ and $z$, with the mean norm increasing smoothly from 2.2 to 2.5 between $z=0$ to 3. This may reflect the more limited set of high redshift objects in our training sample, and suggests that the NF prior is structured in a way that mildly disfavors high redshift solutions relative to the bulk of the sample at low-z. While the impact of the effect is modest in our case (with $|\Delta \ln \pi| \sim 0.7$), such dependencies are important to characterize using a suitable range of training data with known redshifts \citep{spender2}.

%% file: sections/sampling.tex
\subsection{Sampling}
\label{sec:sampling}

While the PAE brings the challenge of inference to a pre-conditioned, reduced-dimension parameter space ($<\mathcal{O}(10)$) relative to the dimension of the data, efficient sampling is nonetheless crucial for inference on a scale relevant for modern galaxy surveys. Furthermore, while we train the NF to transform the autoencoder latent variable distribution into a simpler multivariate Gaussian, there is no guarantee that the posteriors of individual sources are Gaussian in this pre-conditioned space, as is commonly assumed in variational methods. Using an array of ensemble-based annealing techniques, \cite{speagle_ensemble_mcmc_photoz} showed that template-based photo-z likelihood surface of 30-band COSMOS photometry is complex, with non-Gaussian structure and multimodality. 

We deployed and tested two samplers with our PAE model:
\begin{itemize}
    \item the gradient-based Microcanonical Langevin Monte Carlo \citep[MCLMC;][]{robnik_mclmc}, implemented within \texttt{BlackJAX}, a library of samplers designed for composable Bayesian inference \citep{cabezas2024blackjax}, and
    \item the gradient-free, Sequential Monte Carlo sampler implemented in \texttt{pocoMC} \citep{karamanis2022pocomc}.
\end{itemize}
While both methods recover consistent posterior estimates, we determined that our MCLMC implementation is more computationally efficient for large-scale inference, leveraging GPU parallelization and gradient-based updates. We use MCLMC for the results in this work, however we note that \texttt{pocoMC} is suitable in cases with strong posterior multi-modality and/or when only CPU resources are available. 

\subsubsection{Microcanonical Langevin Monte Carlo (MCLMC)}
MCLMC is part of a family of methods that casts the sampling of a target distribution (in our case, the posterior) as an energy-conserving, stochastic process. In particular, given a target density $p(x) \propto e^{-S(x)}$, the distribution is stationary and ergodic under the stochastic differential equation
\begin{equation}
\frac{d}{dt}\begin{bmatrix} x \\ u \end{bmatrix} = \begin{bmatrix} u \\ -P(u)(\nabla S(x)/(d-1)) + \eta P(u) dW\end{bmatrix},
\end{equation}
where $S(x)$ denotes the action, $d$ is the dimension, $u$ is the auxiliary momentum variable (to be distinguished from the NF latent vector), and $\eta$ is the step size.
For high-dimensional problems, MCLMC has been shown to offer several-fold improvements in sampling speed relative to the commonly used Hamiltonian Monte Carlo (HMC) and the No U-turn sampler (NUTS) \citep{bayer23, simon25, sommer25}; however, we find improvements in speed even for the low-dimensional inference problems in this work. For smooth, convex potentials, MCLMC sampling converges exponentially fast, which is advantageous in forecasting computational needs for LSS surveys at scale.

We run four chains per source in order to evaluate convergence. We use three stages of sampling:
\begin{enumerate}
    \item \textit{Burn-in and re-initialization:} We initialize chain positions by drawing from the priors $\mathbf{u}\sim \mathcal{N}(0, \mathbf{I})$ and $z\sim \text{Unif}(z_{\rm min}, z_{\rm max})$ and run MCLMC for 500 steps. For each source, we use the position of the chain with the highest log-likelihood to re-initialize the other chains with significantly poorer likelihood (i.e., with $\Delta \ln \mathcal{L} > 5$), and a small amount of scatter to the new positions. We find this step can help correct for chains that get stuck in erroneous local minima during burn-in.
    \item \textit{MCLMC pre-tuning:} We run MCLMC with pre-tuning to determine the optimal step size $\eta$ and momentum decoherence scale $L$ of each chain. This is run for $\sim 30\%$ the length of the final sampled chains.
    \item \textit{Posterior sampling:} With optimized $\eta$ and $L$ for each chain, we run MCLMC for 2000 steps, discarding the first 1000 samples of each chain when collecting our final posteriors.  
\end{enumerate}

For numerical stability, we sample in log-redshift space rather than absolute redshifts, with the corresponding Jacobian factored into our log-density. This ensures positive definite redshift samples and avoids numerical divergence issues stemming from the fact that MCLMC utilizes continuous trajectories in parameter space that occasionally explore regions $z<0$ without reparameterization.

In Appendix \ref{sec:convergence} we evaluate the redshift convergence properties of our sampling, computing the Gelman-Rubin statistic $\hat{R}$, which quantifies the relative variance across and within chains, as well as the mean autocorrelation length across chains for each source.





\subsection{Computational performance}

In order to perform inference at a scale relevant to modern galaxy surveys, we design an implementation, which we call \texttt{PAESpec}, that is both efficient and leverages the parallel computing capabilities of modern infrastructure. Our testing was performed on A100 GPU nodes at NERSC\footnote{\url{https://www.nersc.gov/}}, each of which is equipped with 40 GB of RAM. 

The computational performance of \texttt{PAESpec} is primarily driven by the cost of the likelihood evaluation, with minimal overhead from function compilations. Our final PAE model performs full forward passes from $\mathbf{u}, z$ to model photometry in $\sim 2$ ms. This is driven, in rank order, by the normalizing flow forward pass (1 ms), redshifting/interpolation\footnote{We note that utilizing shift+interpolate operations for redshifting on a logarithmic wavelength grid offers a considerable improvement in speed performance compared to our baseline, and will be incorporated in future implementations.} (0.5 ms) the decoder forward pass (0.3 ms) and filter convolution (0.1 ms). Using MCLMC, we achieve sampling rates of $\sim 300$ samples per second, and find that gradient evaluations during the sampling process account for only a small portion of the total sampling cost. Our implementation utilizes GPUs and employs just-in-time (JIT) compilation, along with a differentiable forward model, to facilitate MCLMC and other gradient-based algorithms. 

To deploy \texttt{PAESpec} at scale, we leverage parallelization through the distributed computing capabilities of JAX. In particular, we use \texttt{jax.vmap} to vectorize computations across multiple chains and different sources within a single device. This enables us to run approximately 1000 chains simultaneously on a GPU core without appreciable increase in wall clock time. We optionally use \texttt{jax.pmap} to vectorize computations across multiple GPU devices when resources are available. 

%% file: sections/pae_tests.tex
\section{SED Reconstruction}
\label{sec:results}

For the sake of illustration and simplicity, we use a PAE with $n_{\rm latent}=5$ for the results presented in this section, though for our fiducial redshift results we use $n_{\rm latent}=10$ (see Appendix \ref{sec:vary_nlatent} for a more detailed discussion).

\subsection{Reconstruction statistics}
\begin{figure}
    \centering
    \includegraphics[width=0.9\linewidth]{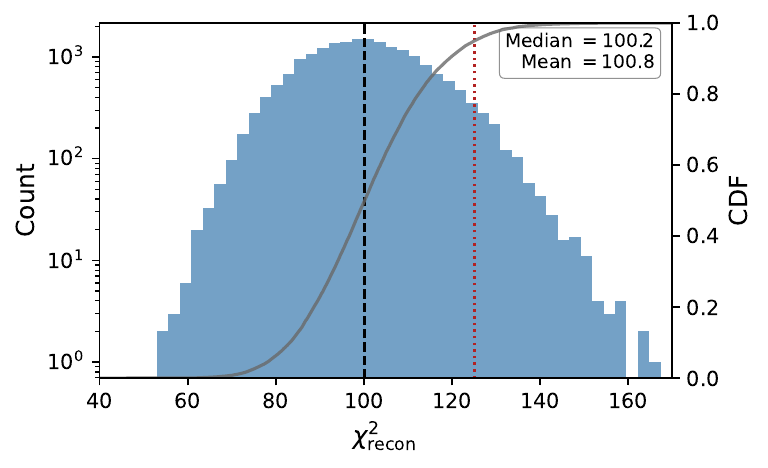}
    \caption{Distribution of PAE reconstruction $\chi^2$, derived from the posterior mean for each source. Also plotted on the twin y-axis is the CDF of the $\chi^2$ distribution. The vertical lines indicate the median (black dashed) and \nth{95} percentile (red dotted) of the $\chi^2$ distribution.}
    \label{fig:chi2_recon}
\end{figure}

We begin in Figure \ref{fig:chi2_recon} by showing the distribution of $\chi^2_{\rm{recon}}$ for 20,000 galaxies in our validation sample. We find a median $\chi^2=100.2$ -- for a six parameter model (five latents + redshift), $\chi^2_{\rm{red}} = 100.2/(102-6)=1.04$, indicating excellent goodness-of-fit. Ninety five percent of sources have $\chi^2_{\rm red}<1.25$, and $\chi^2_{\rm{max}} \sim 170$, suggesting that our PAE is able to effectively reconstruct the validation data.

\subsection{Individual example}
\begin{figure*}
\centering
\includegraphics[width=0.64\linewidth]{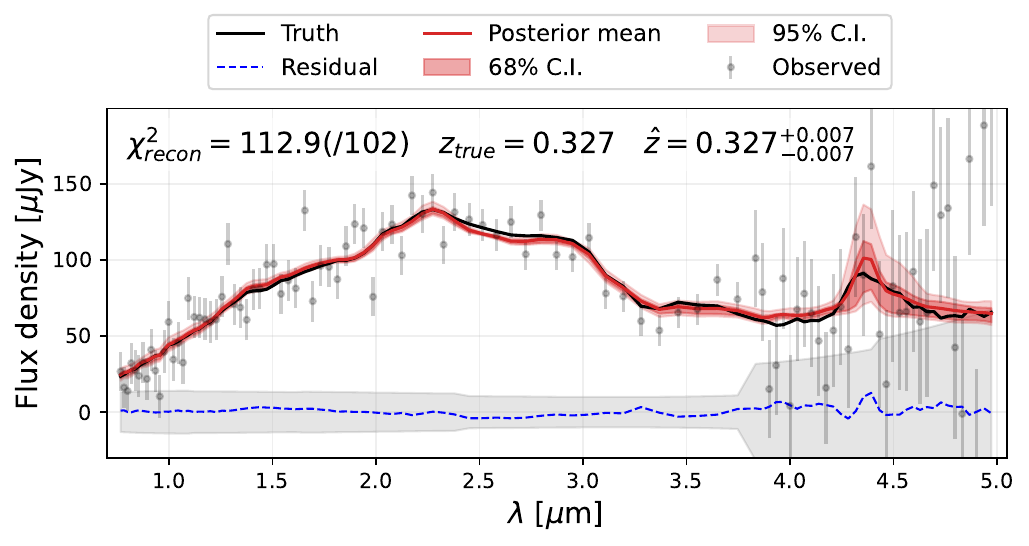}
\includegraphics[width=0.33\linewidth]{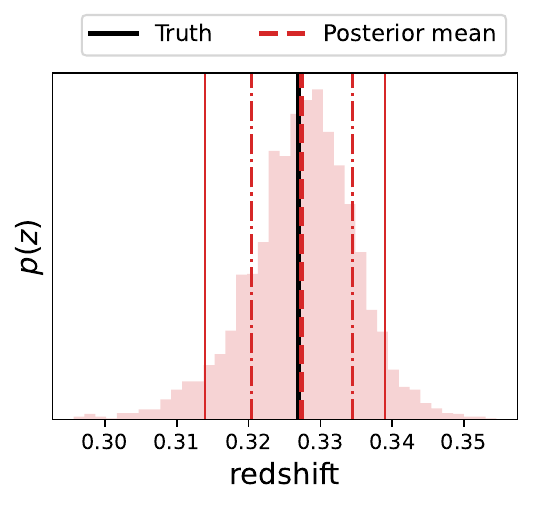}
\caption{Left: Reconstruction results for a bright galaxy, with 68\% and 95\% credible intervals plotted against the input SED (black). Right: Corresponding redshift posterior. Red vertical lines indicate the posterior mean (dashed), 68\% (dot-dashed) and 95\% (solid) credible intervals.}
\label{fig:specpost}
\end{figure*}

\begin{figure}
    \centering
\includegraphics[width=\linewidth]{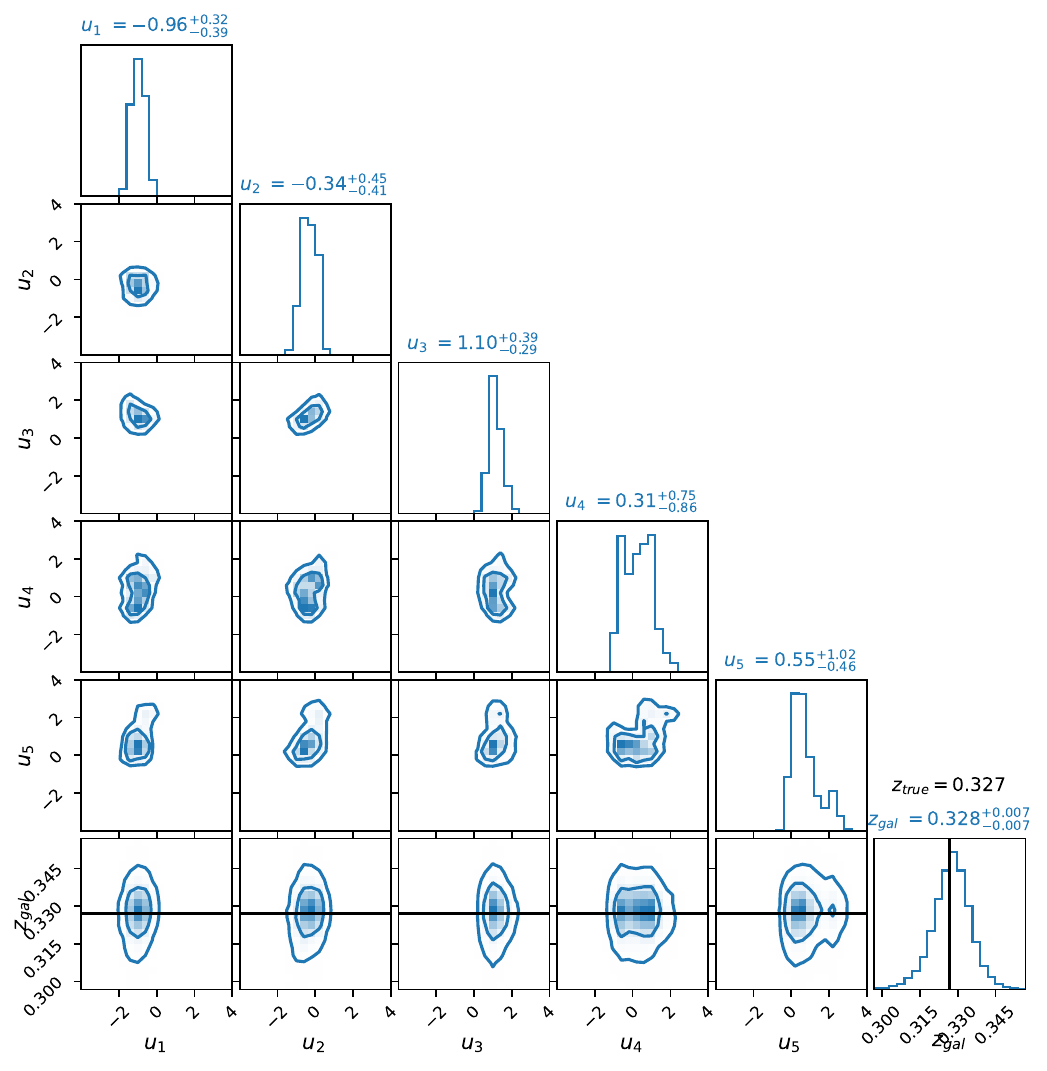}
\caption{Corner plot of the PAE posterior over latent parameters $\lbrace u_i\rbrace$ and redshift $z_{gal}$, for the same source as in Fig. \ref{fig:specpost}. Despite a posterior over latent parameters that is complex, the redshift posterior is well-behaved, providing an accurate estimate of the true redshift.}
\label{fig:latentpost}
\end{figure}

To highlight the quality of our PAE reconstructions, in Figures \ref{fig:specpost} and \ref{fig:latentpost} we show the results for a bright galaxy at $z=0.327$. The reconstruction has decent goodness-of-fit and a posterior predictive distribution with 95\% credible interval that largely covers the true (noiseless) SED. At long wavelengths ($\lambda_{\rm obs} \approx 4.4$ $\mu$m), the fit properly recovers the PAH feature with uncertainties that are consistent with higher noise at those wavelengths. Despite the fact that the full posterior exhibits some non-Gaussian structure and correlations between parameters (Fig. \ref{fig:latentpost}), the posterior over redshift is well-behaved and unbiased. In \S \ref{sec:corr_uz_zerr} we check more rigorously that correlations between redshift and latent parameters do not correspond to source populations with biased redshift estimates.




\subsection{Reconstruction on missing data}

To stress test our model, we run \texttt{PAESpec} on data in which we systematically select random subsets of the complete photometry. For these cases, we utilize the same input data but set the weights of missing bands to zero, ensuring that they do not contribute to the likelihood. 

\begin{figure}
    \centering
    \includegraphics[width=\linewidth]{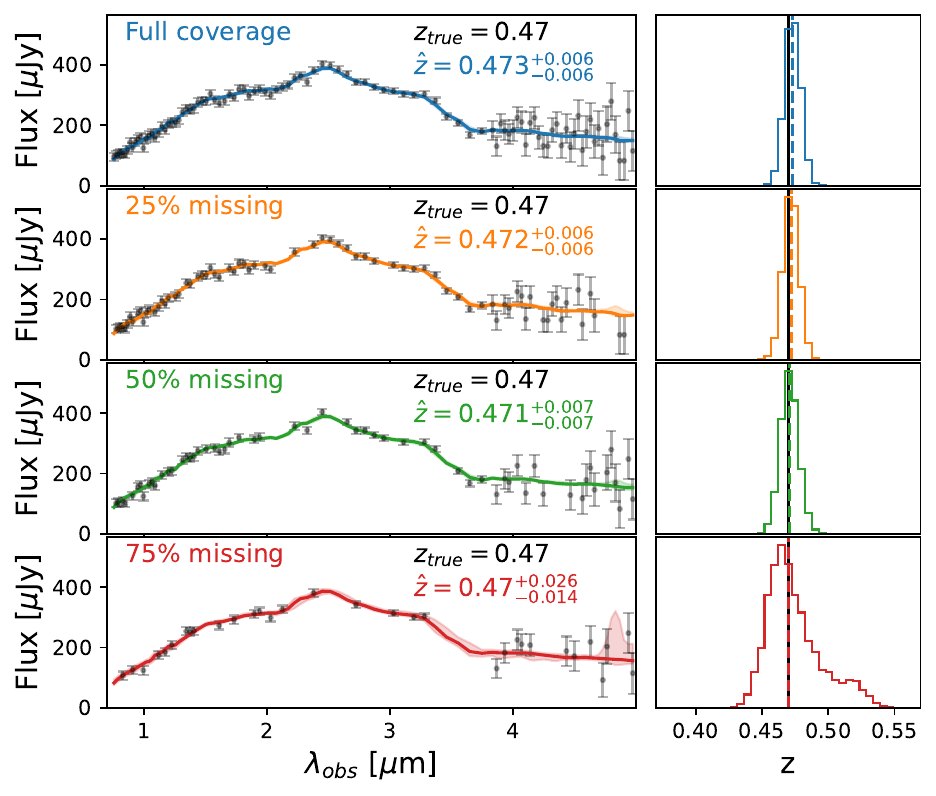}
    \caption{SED reconstruction and redshift estimation for a bright galaxy at $z_{\rm true}=0.470$, with each row showing the results for a subset of the initial flux measurements. For each reconstruction, the shaded region indicates the 95\% credible interval as a function of wavelength. Each right-hand panel shows the resulting redshift PDF and true redshift (black). Our forward modeling is robust to noisy/incomplete data, with commensurate degradation of redshift precision as the fraction of missing data is increased.}
    \label{fig:pae_missing}
\end{figure}

Figure \ref{fig:pae_missing} shows the result of this exercise for a well-detected galaxy at $z_{\rm true}=0.47$. As we progressively reduce the spectral coverage, the redshift uncertainties become larger; however, this degradation is relatively mild, even in scenarios where up to 50\% of the data is missing ($\sigma_z = 0.006$ vs. 0.007 between full coverage and 50\% coverage). The mild degradation may be explained in this case by the detection of spectral features (e.g., the $\lambda_{\rm rest} = 1.6$ $\mu$m bump \citep{sawicki}, the apparent ``shoulder" at $\lambda_{\rm rest}=2.3-2.4$ $\mu$m from CO absorption \citep{gemini_co}), or their combination, which leads to a precise redshift determination despite the partial sampling. In this case our redshift results remain stable and unbiased, even in extreme cases where 75\% of the measurements are excluded.

%% file: sections/redshift_inference.tex
\section{Redshift Estimation Results}
\label{sec:inference}
\subsection{Validation metrics}

We use several metrics to assess the quality of our recovered photo-zs:
\begin{itemize}
    \item Normalized median absolute deviation (NMAD): defined as $1.4\times \text{Median}\left(\frac{|\hat{z}-z_{\rm true}|}{(1+\hat{z})}\right)$.
    \item Median fractional redshift uncertainty: $\tilde{\sigma}_{z/(1+z)}$.
    \item Mean bias of the sample: defined as $\langle (\hat{z}-z_{\rm true})/(1+z_{\rm true})\rangle$.
    \item Redshift Z-score: defined as $Z=(\hat{z}-z_{\rm true})/\hat{\sigma}_z$.
    \item Outlier fraction $\eta_{3\sigma}$, the fraction of sources with $|(\hat{z}-z_{\rm true})/\hat{\sigma}_{z}| > 3$.
    \item Probability integral transform (PIT) distribution: For a range of quantiles ranging from 0 to 1, we calculate the fraction of objects whose $p(z)$ estimates contain the true redshift within the CDF up to a given quantile. For well-calibrated posteriors, the PIT should be uniformly distributed.

\end{itemize}

\subsection{Redshift recovery}

In this section, we present our PAE redshift results applied to mock SPHEREx spectra (with $n_{\rm latent}=10$), and compare performance with the template fitting (TF throughout) code used in \cite{feder23, stickley16}. For the TF implementation, we use the same set of 160 templates used to generate the mock data to compute a model grid that spans redshifts $z=0-3$ with $\delta z=0.002$, enumerates over three dust extinction laws (Prevot, Calzetti and Allen) with extinction spanning $E(B-V)=0-1$ with $\delta E(B-V)=0.05$. This matches the resolution of previous SPHEREx redshift estimation studies. The TF code assumes an uninformative prior over redshift and equal weighting of SED templates. The tests in this work are done on 102-band, ``SPHEREx-only" photometry, whereas the redshift forecasts in \cite{feder23} used additional broadband photometry in the optical and IR from external surveys. 

Because we use the same underlying mock dataset to train and test both methods, we can directly attribute differences in redshift performance to differences in implementation and $p(z)$ estimators. We run both methods on a set of 20,000 $z_{\rm AB}<22.5$ galaxies from our validation set. For our fiducial PAE results, we apply both the normalizing flow prior $p(\mathbf{u})$ and the redshift prior derived in \S \ref{sec:dataset}.

\begin{figure*}[t]
    \centering
    \includegraphics[width=0.9\linewidth]{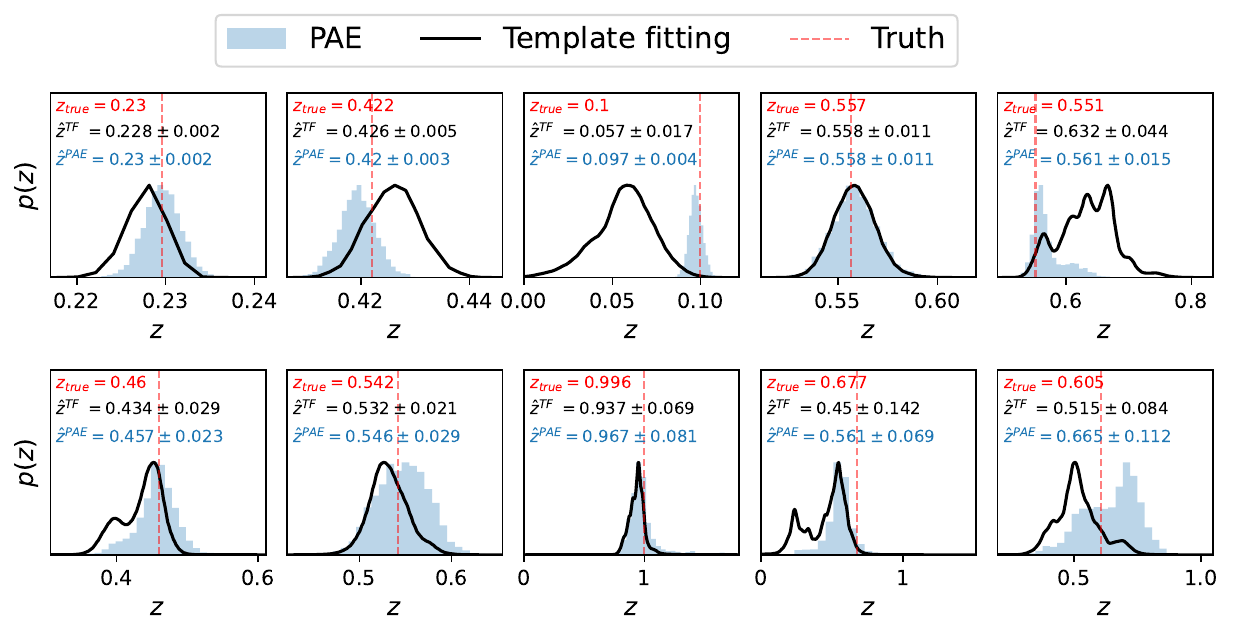}
    \caption{Collection of redshift PDFs from template fitting (black) and PAE (blue), with true redshifts indicated in red. The sources are randomly chosen in bins of increasing $\sigma^{PAE}_{z/1+z}$. While in some cases the estimates from both methods are similar, differences in $p(z)$ shape and peak location are present.}
    \label{fig:compare_zpdfs}
\end{figure*}

In Figure \ref{fig:compare_zpdfs} we show a collection of redshift PDFs $p(z)$ from both methods, for a sample of randomly-selected galaxies with varying redshift precision. While some PAE/TF cases show close agreement (e.g., top row, fourth column), we do find differences. The TF $p(z)$ resolution is limited by the discrete redshift grid (apparent in the highest-precision examples), while MCLMC sampling with the PAE adapts to each individual posterior, yielding smooth, well-sampled PDFs. In some cases, the TF exhibits more complex multi-modal structure (e.g., top row, fifth column and bottom row, fourth column), which may stem from the finite 160-template basis set used to construct the model grid. 

\begin{figure*}
    \centering
    \includegraphics[width=\linewidth]{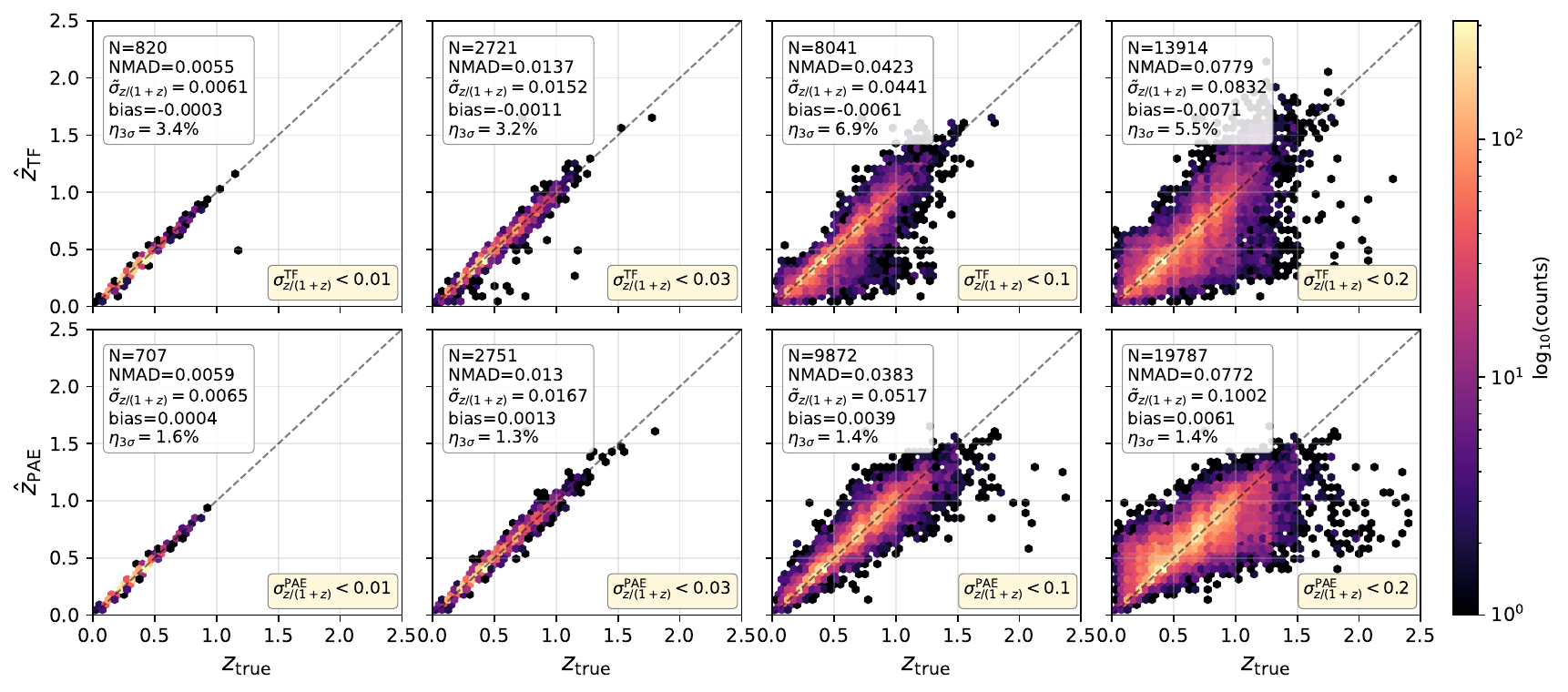}
    \caption{Redshift recovery results obtained with template fitting (TF, top row) and our PAE (bottom), for a $z_{\rm AB}<22.5$ sample. Each column shows a different redshift uncertainty selection, ranging from $\sigma_{z/(1+z)}<0.01$ (leftmost column) to $<0.2$ (rightmost column).}
    \label{fig:compare_tf_pae_zinzout}
\end{figure*}

In Figure \ref{fig:compare_tf_pae_zinzout} we show TF and PAE redshift recovery for the total sample, binned by four redshift uncertainty thresholds ($\sigma_{z/(1+z)}<0.01, 0.03, 0.1$ and 0.2). For the highest-precision selection, $\sigma_{z/(1+z)}<0.01$, the PAE recovers 13\% fewer sources than TF ($N=720$ vs. 820), but recovers similar NMAD (0.0057 vs. 0.0055), a slightly lower $3\sigma$ outlier fraction (2.4\% vs. 3.4\%), and comparable low bias (0.0008 vs. -0.0003). Relaxing the precision threshold to $\sigma_{z/(1+z)}<0.1$, we find that the PAE recovers $\sim 25\%$ more sources than TF, while maintaining small mean biases ($<1\%$), comparable NMAD, and substantially lower 3$\sigma$ outlier fraction (1.6-2.4\% for the PAE vs. 3.2-6.9\% for TF).

For our broadest selection, $\sigma_{z/(1+z)}<0.2$, the PAE recovers nearly 40\% more sources than TF; however, we also see clear redshift-dependent biases of opposite sign canceling in the mean. The distribution of recovered redshifts driving this bias closely matches our prior $p_{\rm BPZ}(z)$ (c.f. Fig. \ref{fig:nzprior}), suggesting the bias is driven by low-SNR, prior-dominated objects in the $\sigma_{z/(1+z)}<0.2$ sample.

\subsubsection{Correlation of redshift errors}
\label{sec:compare_redshift_errors}
\begin{figure}
    \centering
    \includegraphics[width=0.9\linewidth]{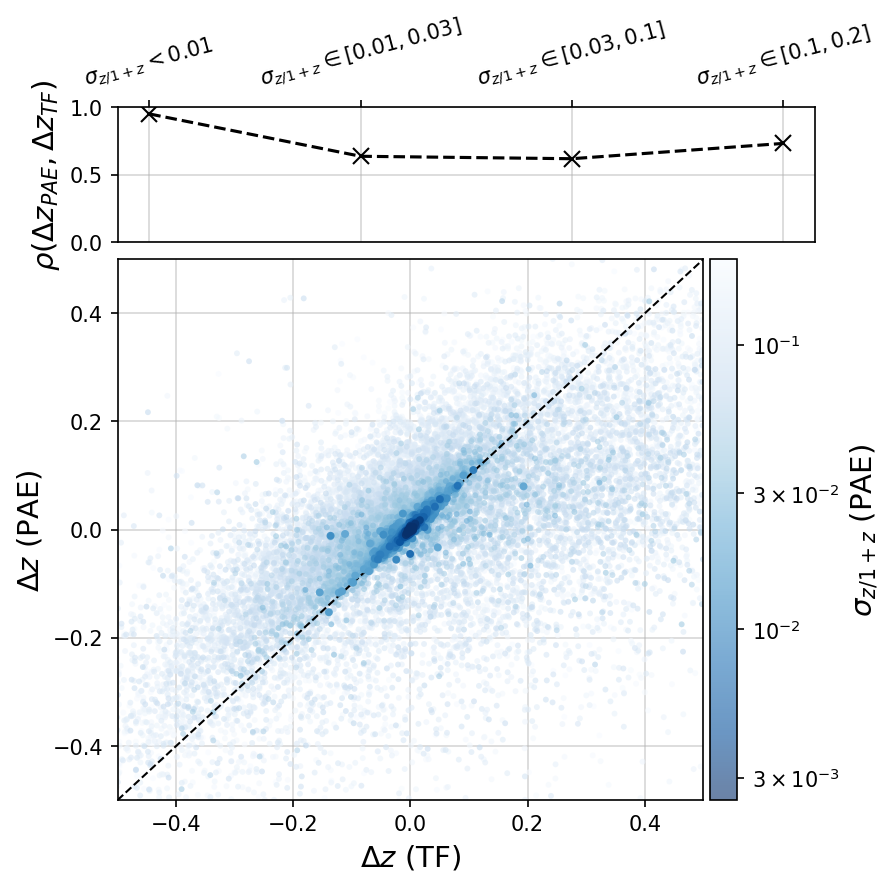}
    \caption{Correlation of redshift errors (i.e., $\Delta z = \hat{z}-z_{true}$). The main panel shows the scatter plot of redshift errors from template fitting (x-axis) and the PAE (y-axis), colored by the PAE fractional redshift uncertainty. The top panel shows the Pearson correlation coefficient of errors between methods, binned by $\sigma_{z/(1+z)}^{\rm PAE}$.}
    \label{fig:zerr_corr}
\end{figure}

We compare the redshift errors $\Delta z = \hat{z}-z_{\rm true}$ from both methods in Figure \ref{fig:zerr_corr}. Importantly, both TF and the PAE results are paired to the same data realizations. It is visually clear that the redshift errors are correlated with each other, which we quantify in bins of $\sigma_{z/(1+z)}$ using the Pearson correlation coefficient. For our highest-precision bin, $\sigma_{z/(1+z)}<0.01$, $\rho$ is nearly unity, which is to be expected in the likelihood-dominated limit. The results in our lower-precision bins exhibit slightly lower correlation, $\rho \sim 0.6-0.75$, which reflects the difference in modeling approaches and assumed priors between methods.  

\subsection{PAE-TF error diagnostics}
\label{sec:pae_tf_errors}
\begin{figure*}[t]
    \centering
    \includegraphics[width=\linewidth]{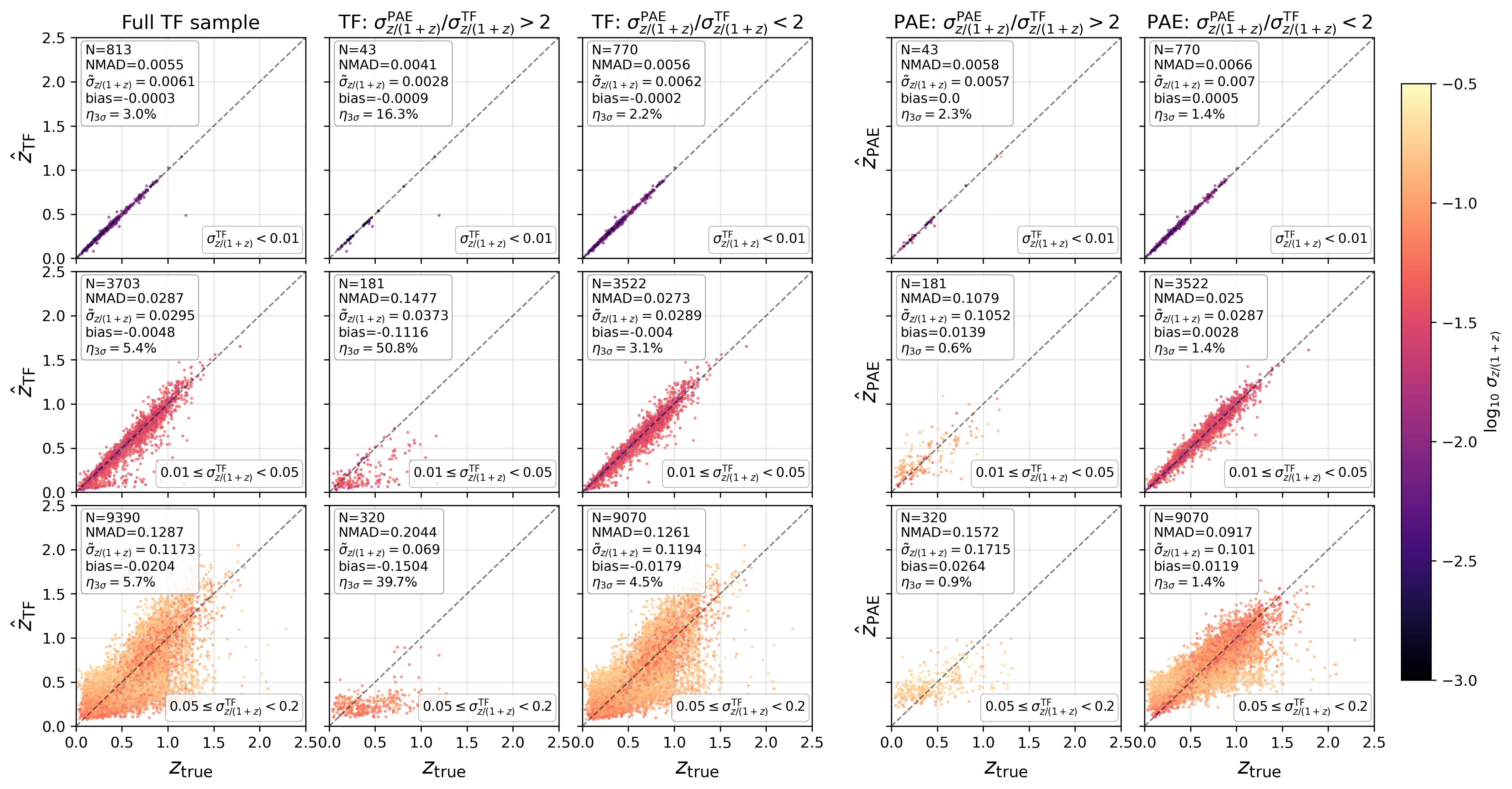}
    \caption{TF redshift recovery results for the full sample (first column) and for sub-samples defined by the PAE + TF selection discussed in the text (second, third columns). The two rightmost panels show the corresponding PAE results with the same selections. Each row shows samples selected by TF redshift uncertainty.}
    \label{fig:compare_tf_pae_with_paesel}
\end{figure*}

While the similarity between PAE and TF estimates helps validate the general consistency of the methods, we find that the differences also provide useful diagnostic power in identifying problematic sub-samples. In this section, we consider a simple cut in which we flag objects with $\hat{\sigma}_{z/(1+z)}^{\rm PAE}/\hat{\sigma}_{z/(1+z)}^{\rm TF}>2$, i.e., sources for which the PAE posterior is considerably broader than the TF estimate. We use PAE results with the BPZ redshift prior applied, which primarily impacts the interpretation of the cut for the lower-precision bins. In Figure \ref{fig:compare_tf_pae_with_paesel}, we show the recovered TF and PAE results for three TF precision bins, splitting the full sample into flagged and unflagged sources.

For all three precision bins, we find that flagged sources are overwhelmingly outliers in the TF recovery. For $\sigma_{z/(1+z)}<0.01$, the flagged sample ($N=43$) has an outlier fraction of 16.3\%, more than 5$\times$ that of the full sample (3.0\%), while the unflagged sample ($N=770$) has reduced outlier fraction (2.2\%) and bias. This effect is even more pronounced in lower-precision bins -- flagged sources in the $0.01\leq \sigma_{z/(1+z)}^{\rm TF} < 0.05$ and $0.05\leq \sigma_{z/(1+z)}^{\rm TF} < 0.2$ have outlier fractions of 50.8\% and 39.7\% respectively, compared with 5.4\% and 5.7\% for the full TF samples. We find similarly stark differences in the mean TF biases for flagged samples, which are between $4-20\times$ that of the unflagged samples depending on the $\sigma_{z/(1+z)}$ bin. 

In contrast, the PAE estimates for both flagged and unflagged sources (fourth column) are well behaved, with minimal biases and outlier fractions ranging from 0.6-2.3\%. We note that the flagged population exhibits systematically worse PAE sampling convergence than unflagged sources in the same TF precision bin. For sources with $\sigma_{z/(1+z)}^{\rm TF}<0.01$, 77\% of flagged sources have $\hat{R}<1.1$ with median $\hat{R}=1.033$, compared to 97\% and median $\hat{R}=1.01$ for unflagged sources in the same bin. For the $\sigma_{z/(1+z)}^{\rm TF} < 0.05$ bin, the corresponding fractions are 84\% and 95\% with medians of 1.025 and 1.013 respectively. The degradation is concentrated in a tail of poorly-converged sources rather than being a feature of the full flagged population. We interpret the poorer convergence as reflecting more complex posterior geometry in the flagged population of sources, which may contribute to the larger $\sigma_{z/(1+z)}^{\rm PAE}$ driving our selection.

While a dedicated selection that balances redshift purity and completeness in the TF sample is beyond the scope of this work, these results highlight the value of comparing TF and PAE $p(z)$ estimates for identifying and removing TF outliers, and for interpreting redshift errors more generally.   

\subsubsection{Correlation of PAE errors with NF latents}
\label{sec:corr_uz_zerr}
\begin{figure}
    \centering
    \includegraphics[width=0.9\linewidth]{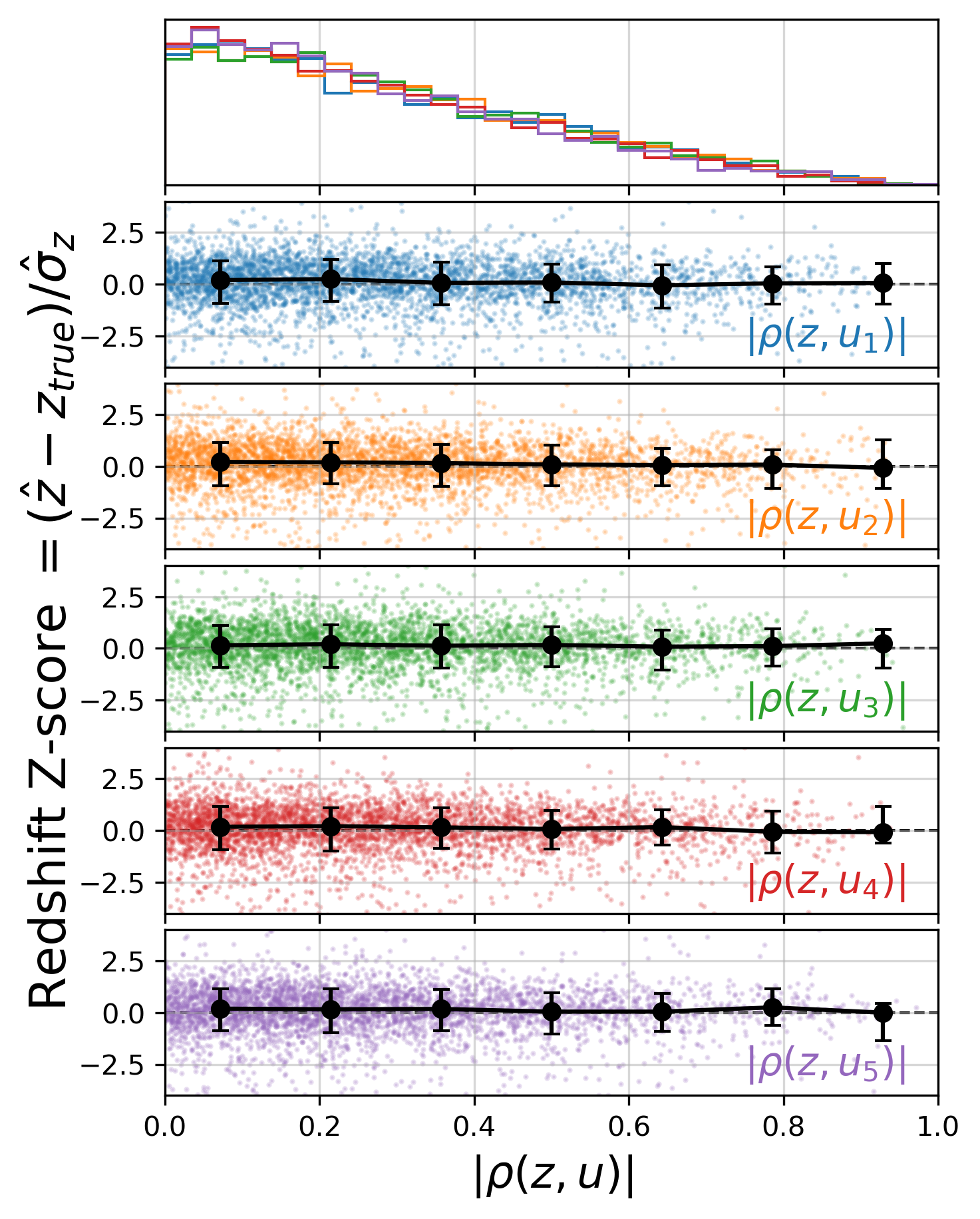}
    \caption{Scatter plots showing the distribution of PAE redshift Z-scores as a function of absolute correlation between redshift and normalizing flow latents $\lbrace u_i\rbrace$, which are computed from each PAE posterior. We show the $n_{\rm latent}=5$ case here for simplicity. The black errorbars indicate the 1$\sigma$ dispersion of the sample, binned in $\rho$. This shows that the redshift errors, normalized by recovered uncertainty, are not impacted by correlations or degeneracies in the posterior parameter space.}
    \label{fig:corr_zu_vs_zscore}
\end{figure}

Our PAE is designed to de-couple variations in redshift from the intrinsic SED variations learned by our rest-frame autoencoder. We perform an empirical check using the measured Pearson correlation coefficient $\rho$ between the PAE redshift and NF latents $\lbrace u_i \rbrace$, which is derived from all posterior samples per source. To check for trends with $\rho$ and potential redshift biases, in Fig. \ref{fig:corr_zu_vs_zscore} we show the distribution of measured correlation strength across latent variables, as well as the distribution of redshift Z-scores as a function of $|\rho|$. For compactness we show the case for $n_{\rm latent}=5$, however the conclusions are the same for all choices of $n_{\rm latent}$. In particular, we do not see any significant biases as a function of $\rho(z,u)$, confirming that, even for cases with stronger correlation between redshift and rest-frame ``nuisance" parameters, the marginalized redshift estimates are unbiased.

\subsection{Impact of priors}
\label{sec:prior_impact}

A feature of our Bayesian framework is the ability to easily specify and vary the priors entering the PAE posterior. Here we assess the sensitivity of our redshift results to two prior choices: the normalizing flow (NF) prior on latent parameters $\pi(\mathbf{u})$, and the redshift distribution prior $\pi(z)$. We test four configurations with/without a NF prior and using uniform $p(z)$ vs. $p_{\rm BPZ}(z)$. We summarize the results across all prior combinations in Table~\ref{tab:compare_priors} and compare redshift errors, redshift Z-scores, and PIT distributions across methods and prior configurations in Fig.~\ref{fig:zpdf_calib}, binned by $\sigma_{z/(1+z)}$. To quantify the distance between the empirical and ideal (uniform) PIT distributions we compute the Kolmogorov-Smirnov (KS) statistic for each case.

\begin{table*}[t]
\centering

\begin{tabular}{c|c|c|c|c|c|c}
Method & Prior choice & $N_{\rm gal}$ & $\langle \Delta z/(1+z) \rangle$ & $\langle \hat{\sigma}_{z/1+z}\rangle $ & NMAD & $\eta_{3\sigma}$ [\%] \\
 \hline
TF & Uniform over $z$ and model params & 1805 (13914) & -0.0005 (-0.007) & 0.0108 (0.088) & 0.0097 (0.078) & 2.8 (5.5) \\ \hline
PAE & No prior on $\pmb{u}$; $z\sim \text{Unif}[0, 3]$ & 1284 (8663) & 0.0014 (0.015) & 0.0111 (0.092) & 0.0091 (0.066) & 2.8 (6.9) \\
 & $\pmb{u}\sim \mathcal{N}(0, \pmb{I})$; $z\sim \text{Unif}[0, 3]$ & 1691 (10529) & 0.0007 (0.016) & 0.0114 (0.087) & 0.0092 (0.062) & 1.8 (4.9) \\
 & No prior on $\pmb{u}$; $z\sim p_{\rm{BPZ}}(z)$ & 1229 (17388) & 0.0015 (-0.003) & 0.0113 (0.101) & 0.0103 (0.101) & 3.4 (4.2) \\
 & $\pmb{u}\sim \mathcal{N}(0, \pmb{I})$, $z\sim p_{\rm{BPZ}}(z)$ & 1706 (19787) & 0.0008 (0.006) & 0.0114 (0.100) & 0.0092 (0.077) & 1.4 (1.4) \\
\end{tabular}

\caption{Summary of redshift estimation results, showing the statistics for samples with $\sigma_{z/(1+z)}<0.02$ (and for $<0.2$ in parentheses). We show the results from template fitting as well as the PAE under different prior choices corresponding to Fig. \ref{fig:zpdf_calib}.}
\label{tab:compare_priors}
\end{table*}

\subsubsection{Sensitivity to NF and redshift priors}

\begin{figure*}[t]
    \centering
    \includegraphics[width=0.85\linewidth]{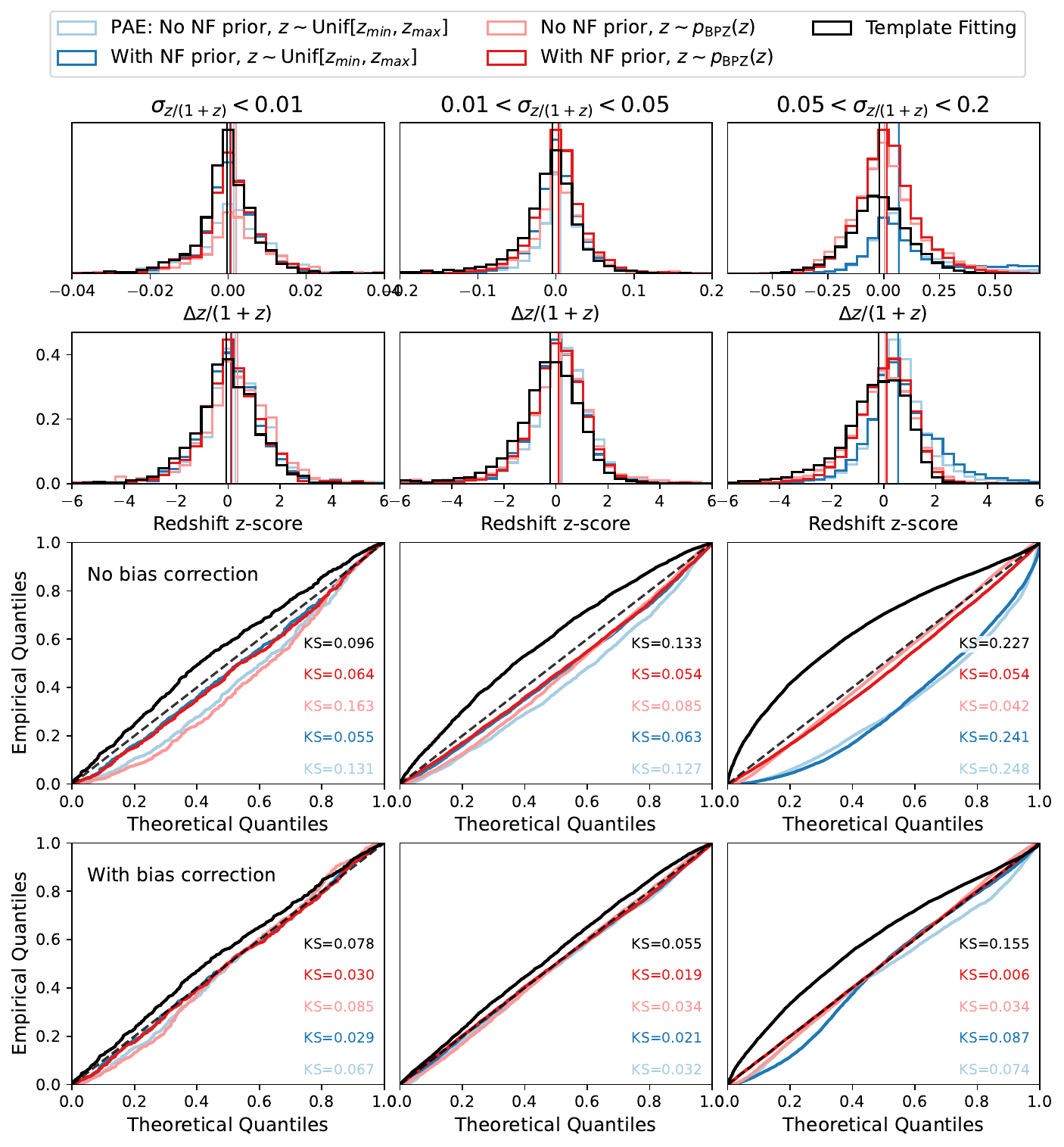}
    \caption{Comparison of redshift results between template fitting (black) and our PAE under different prior configurations (colored histograms). Each column corresponds to samples selected by different fractional redshift uncertainty $\sigma_{z/(1+z)}$. In the first and second rows, we plot the fractional redshift errors and redshift Z-scores, respectively. In the third row, we show empirical coverage plots to assess the calibration of redshift PDFs, while in the last row we show the same coverage plots but after de-biasing the PDFs according to the mean redshift bias of each selected sample. We report the KS-statistic of all PIT distributions (relative to a uniform distribution) corresponding to the different prior configurations/sub-samples in the bottom right corner of each panel.}
    \label{fig:zpdf_calib}
\end{figure*}

Across all configurations in Table \ref{tab:compare_priors} and Fig. \ref{fig:zpdf_calib}, the NF prior $u\sim \mathcal{N}(0, \mathbf{I})$ acts primarily as a regularizer on the SED latent space manifold. At fixed redshift-precision cuts, the NF prior increases the number of recovered sources (by $25-35\%$ for the high-precision sample and by $5-10\%$ for low-precision) and reduces the $3\sigma$ outlier fraction by roughly a factor of two for all selections except the low-precision sample with uniform redshift prior. These improvements are consistent with the NF prior discouraging chains from exploring low-density regions of latent space that produce formally acceptable $\chi^2$ values but implausible SEDs. In Appendix~\ref{sec:convergence} we find that the NF prior also substantially improves sampling behavior -- chain mixing and convergence as quantified using the Gelman-Rubin statistic $\hat{R}$ and chain autocorrelation lengths are small, whereas chains run without the NF prior have a tail extending to higher $\hat{R}$ and $\tau$. For posterior calibration, the NF prior tends to improve coverage for the bright/high-precision samples, as evidenced by the lower KS values in Fig. \ref{fig:zpdf_calib}.

The impact of the redshift prior is strongest in the low-SNR/low-precision regime. Under a uniform $p(z)$, both PAE configurations recover 30-40\% fewer sources (compared to TF) and show clear miscalibration in the broadest bin ($0.05<\sigma_{z/(1+z)}<0.2$): the fractional error and Z-score distributions contain extended tails, and the PIT curves in Fig. \ref{fig:zpdf_calib} deviate strongly from a uniform distribution, reflecting a population of incorrect high-redshift solutions. Selecting on these sources with $0.1<\sigma_{z/(1+z)}<0.2$, we find that the sources with erroneous high-$z$ solutions have a median SNR of 16.7, and tend to have redder colors ($\langle z-W1 \rangle = 2.5$) than the full low-precision sample ($\langle z-W1 \rangle = 1.6$).

Including the informative $p_{\rm BPZ}(z)$ prior effectively mitigates this failure mode by down-weighting high-$z$ solutions that are inconsistent with the expected population-level $N(z)$. Empirically, this yields a large increase in the number of recovered sources with $\sigma_{z/(1+z)}<0.2$ and produces substantially better-calibrated posteriors. The effectiveness of an informative $\pi(z)$ prior depends on how accurately $p(z)$ can be specified for the desired application, along with its robustness under domain shifts between training/calibration fields and the survey sample.

To separate bias from uncertainty misestimation in our coverage plots, we additionally apply an idealized ``clustering redshift-style" calibration: within each sub-sample, we estimate the mean fractional redshift bias and use it to de-bias the per-object PDFs before recomputing the empirical coverage. This correction has a minimal impact on the highest-precision bins. However, in the lower-precision bins, de-biasing moves the PIT curves noticeably closer to ideal coverage, suggesting that a non-trivial source of the mis-calibration is attributable to a coherent mean offset. Nonetheless, residual departures from uniformity remain most clearly for the TF and uniform $p(z)$ cases, which is consistent with the broader tails in the redshift error distribution. These effects are relevant for applications that directly select samples based on reported $\sigma_{z/(1+z)}$.

Taken in full, these prior sensitivities underscore important differences between methods: TF reports a profile likelihood over a discrete, constrained nuisance parameter grid, which introduces implicit regularization (both from the finite template set and per-redshift optimization that compresses directions in the nuisance parameter space). The PAE, by contrast, marginalizes over a continuous latent SED manifold, which reduces discretization errors but also exposes a broader space of models consistent with the same data.

\subsubsection{Comparison of profile likelihoods}
\label{sec:profile}
Profile likelihoods allow us to isolate one specific source of difference between the PAE and TF: the shape of the likelihood surface $\mathcal{L}(\mathbf{f}|z)$ itself. In \S \ref{sec:compare_redshift_errors} we identified a population of flagged sources for which the PAE posterior is substantially broader than the TF estimate -- here, we use those sources to directly probe whether the disagreement arises from the underlying likelihood surface or from marginalization over a larger posterior volume. These two mechanisms have different implications for interpreting the PAE model.

Standard $\chi^2$ minimization codes such as \texttt{LePhare} and the SPHEREx in-house code \citep[detailed in][]{stickley16} derive redshift estimates through the profile likelihood: for each redshift step (typically on a discrete grid), nuisance parameters $\psi$ (template index, dust law/strength, and amplitude) are optimized and the maximum likelihood is recorded, 
\begin{equation}
\mathcal{L}_{\rm prof}(z) \equiv \max_{\psi}\,\mathcal{L}(\mathbf f\,|\,z,\psi),
\label{eq:lprof_tf}
\end{equation}
with a redshift PDF obtained from $\mathcal{L}_{\rm prof}(z)$. Profiling corresponds to an implicit delta-function treatment of $\psi$ at its best-fit value over a constrained parameter space. The success of the profile likelihood, however, hinges on a number of conditions: the nuisance parameters $\psi$ are well-constrained by the data; the likelihood is close to Gaussian-distributed with respect to $\psi$; the target parameter (in our case, redshift $z$) is not strongly correlated with $\psi$; and the underlying model is well-specified, i.e., the model adequately describes the data. Whether these conditions are satisfied is unclear in general -- correlations between redshift, dust attenuation and SED type are well documented in photo-z studies \citep{benitez_bpz, eazy}, and it has been shown that photo-z likelihoods can have strong non-Gaussian and multi-modal structure \citep{speagle_ensemble_mcmc_photoz}. In the low signal-to-noise limit, nuisance parameters may be poorly constrained, making the profile likelihoods flatter and/or more multi-modal. These non-idealities can potentially lead to underestimated uncertainties, incomplete coverage of multi-modal solutions and biased redshift estimates. 

In contrast, our PAE performs full marginalization over latent parameters $\mathbf{u}$ to obtain redshift posteriors,
\begin{equation}
    p(z) = \int_{\mathbf{u}} \mathcal{L}(\mathbf{f}|z, \mathbf{u}) \pi(\mathbf{u}) \pi(z) d\mathbf{u}.
\end{equation}
To compare with TF on similar footing, we compute an analogous PAE profile likelihood over latent parameters,
\begin{equation}
\mathcal{L}^{\rm (PAE)}_{\rm prof}(z) \equiv \max_{\mathbf{u}}\,\mathcal{L}(\mathbf f\,|z, \mathbf{u}),
\label{eq:lprof_pae}
\end{equation}
which isolates the best-fit likelihood independent of marginalization effects. We compute $\mathcal{L}^{\rm (PAE)}_{\rm prof}(z)$ on a grid of 300 fixed redshifts between $0<z<3$ using limited-memory BFGS (L-BFGS) to minimize the negative log-likelihood over $u$ at each redshift, holding $z$ fixed in the forward model. We run 100 L-BFGS iterations and perform five restarts to mitigate failed convergence. 

\begin{figure*}
    \centering
    \includegraphics[width=0.49\linewidth]{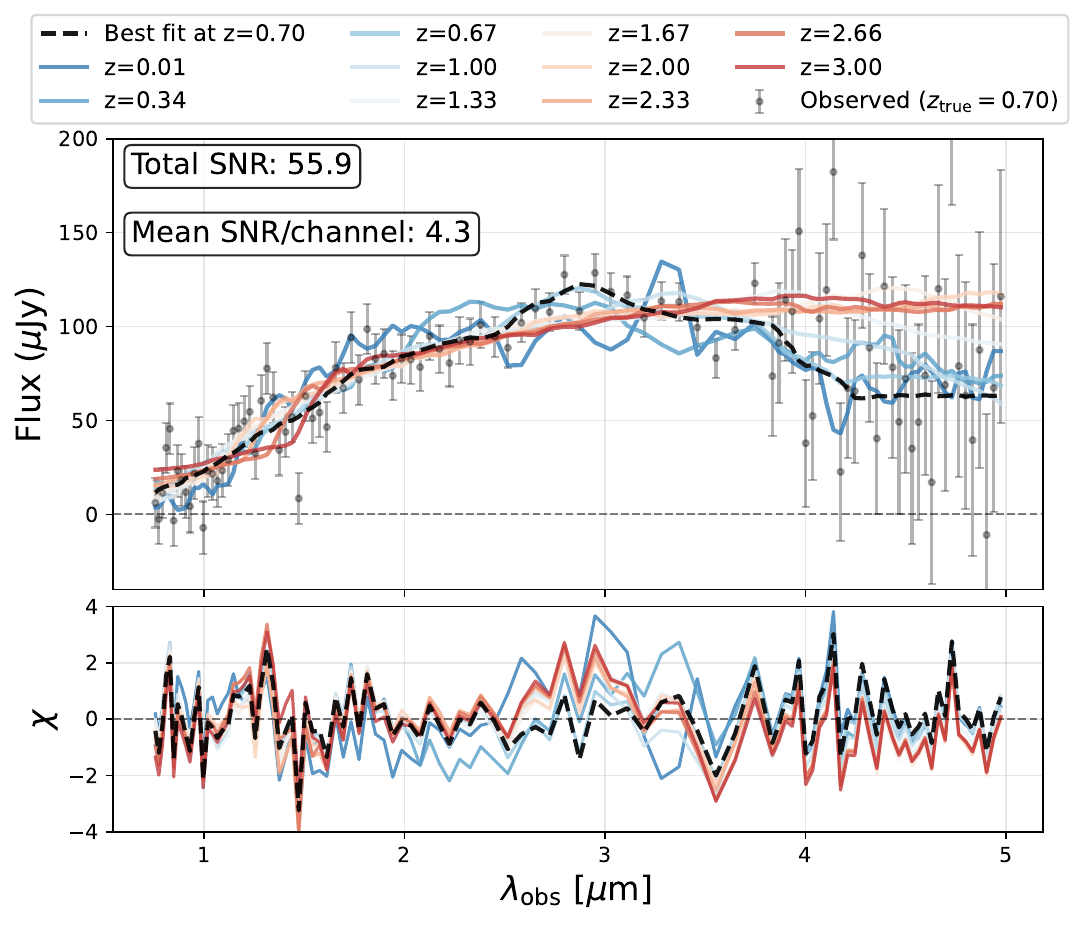}\includegraphics[width=0.5\linewidth]{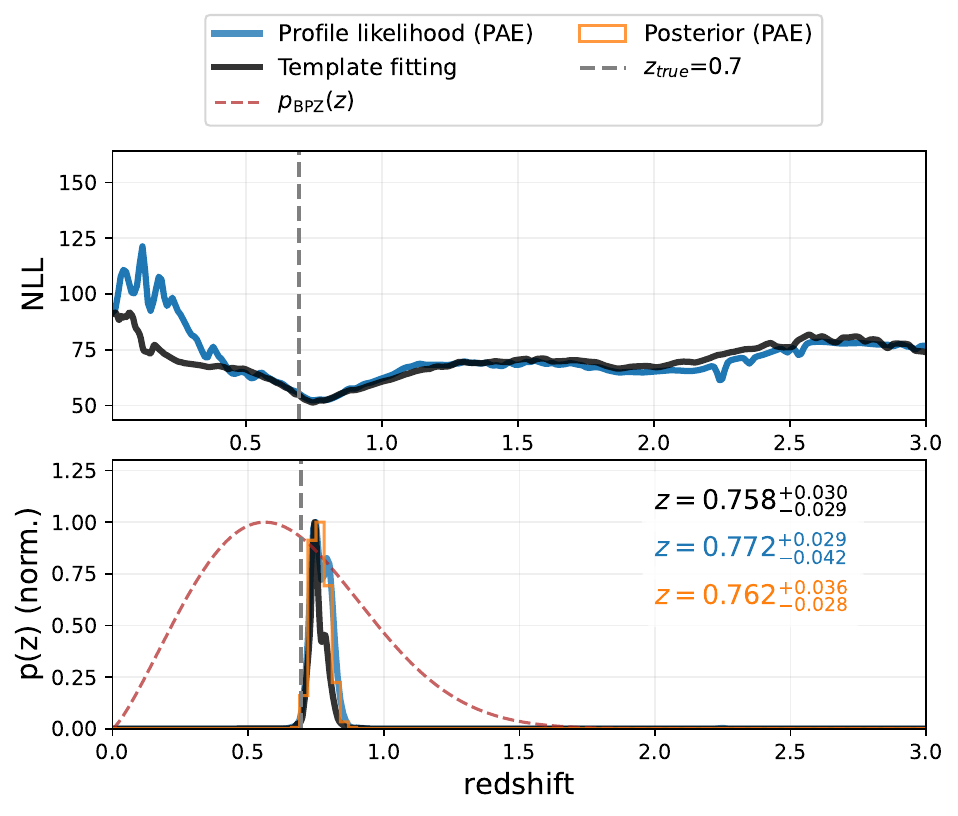}
    \includegraphics[width=0.49\linewidth]{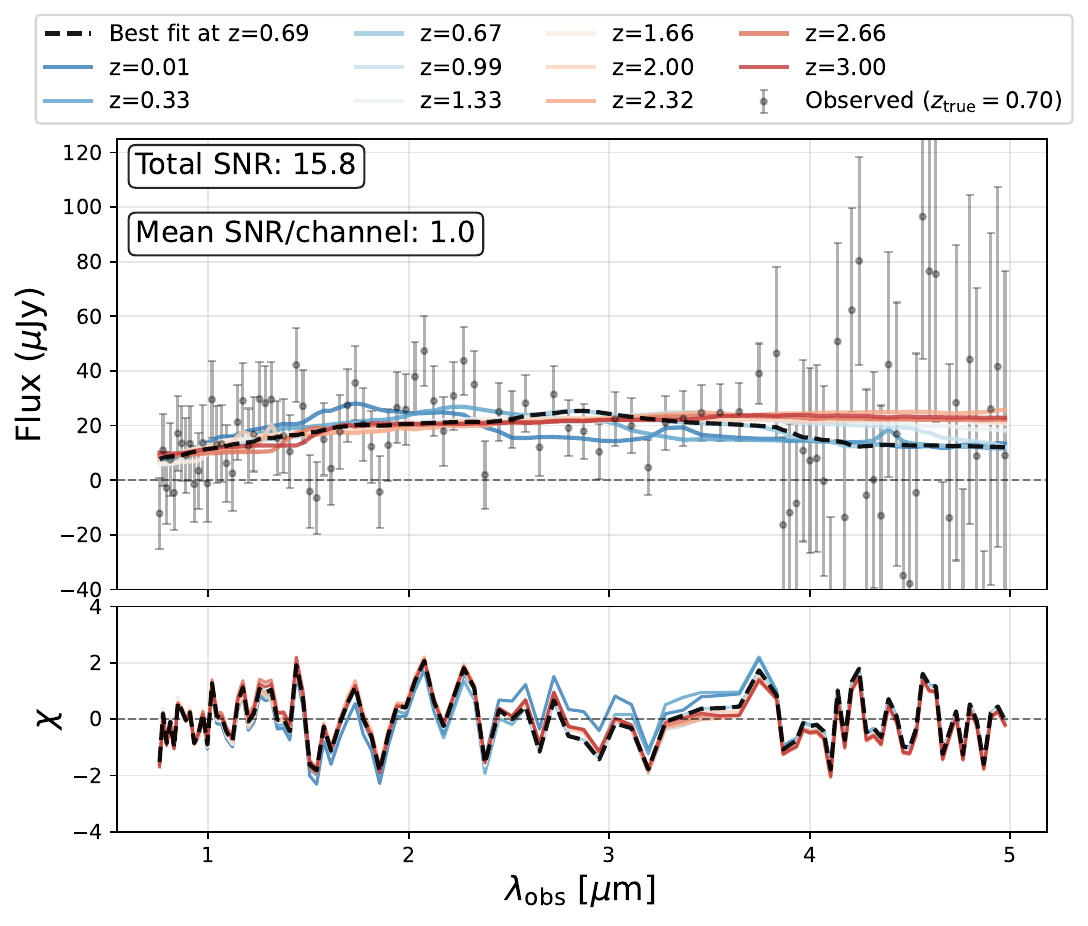}\includegraphics[width=0.5\linewidth]{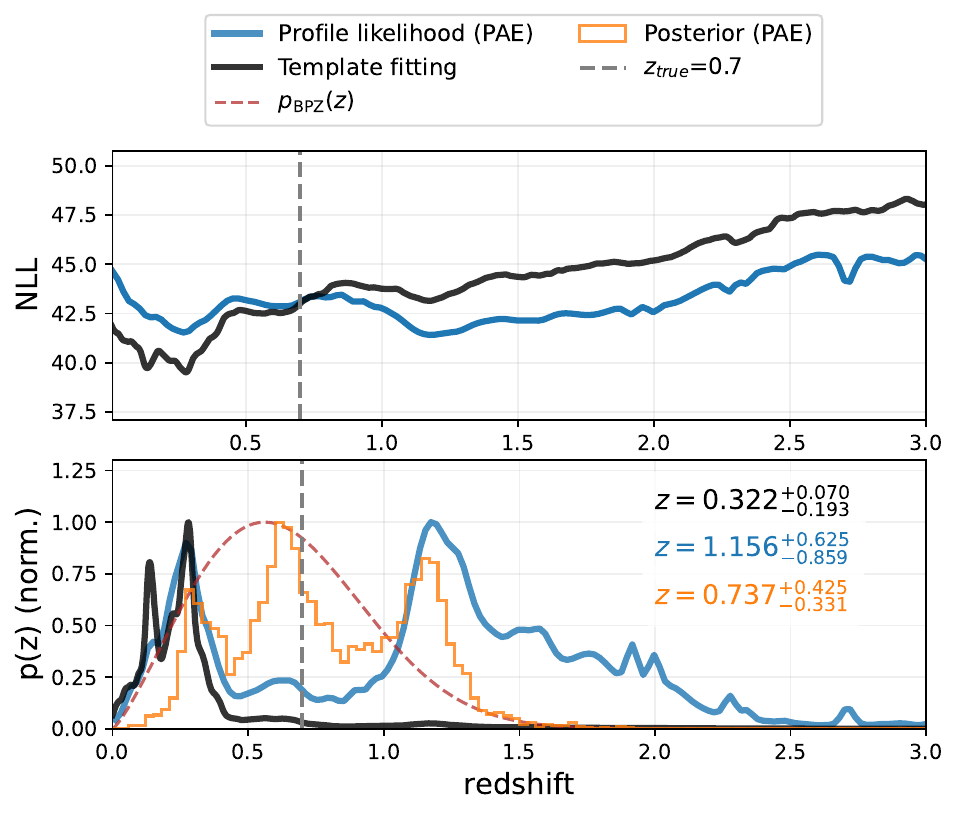}
    \caption{\textbf{Left}: Best-fit models at selected redshifts optimized with LBFGS compared against simulated SPHEREx spectra. We quote the total signal-to-noise ratio (SNR) and the average SNR per channel. Each dashed curve corresponds to the best-fit model closest in redshift to $z_{\rm true}$. The bottom panels show the corresponding normalized residuals $\chi = (f_{\rm obs}-\hat{f})/\sigma_f$. \textbf{Right}: Comparison of profile likelihoods obtained through template fitting (black), our PAE model minimized with LBFGS (blue) and PAE posteriors using MCLMC (orange). The top panels show the negative log-likelihoods recovered from each method while in the bottom panels we show the corresponding photo-z PDFs.}
    \label{fig:pl_comparisons}
\end{figure*}

In Figure \ref{fig:pl_comparisons} we show results for two illustrative cases. We start with a well-detected galaxy at $z_{\rm true}=0.70$ with SNR$=55.9$. Here, the PAE and TF profile likelihoods are in close agreement across the majority of the redshift range, with both exhibiting a sharp minimum near the true redshift. The corresponding $p(z)$ estimates from TF profiling ($\hat{z}=0.758^{+0.030}_{-0.029}$), PAE profiling ($\hat{z}=0.772^{+0.029}_{-0.042}$) and PAE posteriors from MCLMC ($\hat{z}=0.762^{+0.036}_{-0.028}$) are highly consistent with one another, validating that in the high-SNR regime, both methods probe a similar likelihood surface. The differences between the global minimum and local minima/plateaus are large ($\Delta \log \mathcal{L} > 20$). For redshifts $z<0.5$, the TF solutions achieve a lower negative log-likelihood (NLL) than the PAE, however these deviations are far from the true solution and thus do not affect the final results. The best-fit SEDs at selected redshifts from the PAE profiling procedure (Fig. \ref{fig:pl_comparisons}) further illustrate this, with clear differences between the optimized SEDs and significant deviation in model $\chi$ residuals in the range $\lambda_{\rm obs} =2-4$ $\mu$m, highlighting the importance of this wavelength range for redshift determination with SPHEREx. 

For lower-SNR cases, the picture changes substantially. The second example in Fig. \ref{fig:pl_comparisons} shows a fainter galaxy at $z_{\rm true}=0.7$ with total SNR $=15.8$, which satisfies the criterion $\sigma_{z/(1+z)}^{\rm PAE}/\sigma_{z/(1+z)}^{\rm TF}>2$ identified in \S \ref{sec:compare_redshift_errors} as an efficient selection for TF redshift outliers, from which we further sub-select sources with $0.05<\sigma_{z/(1+z)}^{\rm TF}<0.1$. In contrast to the first case, the NLL surfaces from both methods are shallow, with a total $\Delta \log \mathcal{L} \approx 8$ across $0<z<3$ for TF and $\Delta \log \mathcal{L} \approx 3$ for the PAE. The best-fit SEDs reinforce this picture, showing nearly indistinguishable shape in the observed frame. The PAE profile covers the same modes as the TF, it reaches consistently lower NLL for $z>1.0$, finding multiple local minima and a correspondingly broader PDF. For this low-SNR case, the TF estimate is confident but incorrect, while the PAE covers the true solution but is uninformative on its own. In the PAE posterior, the redshift prior $\pi_{\rm BPZ}(z)$ disfavors the high-$z$ likelihood support and re-weights the existing NLL minima, leading to an estimate that is broader than from TF but centered on the true redshift ($\hat{z}_{\rm PAE} = 0.737^{+0.425}_{-0.331}$ vs. $z_{\rm true}=0.70$).  

Notably, we find the same qualitative behavior across all ten flagged sources for which we computed profile likelihoods. In each case, TF achieves lower NLL than the PAE at $z\lesssim 0.3-0.5$, while the PAE reaches lower NLL for $z\gtrsim 0.5$. The TF preferred redshifts cluster at $z\sim 0.1-0.3$ for all flagged sources independent of true redshift, consistent with a spurious low-$z$ minimum from the TF's discrete template grid at low SNR, rather than a genuine SED preference. The difference in NLL at low-$z$ may also be affected by reconstruction error in the AE training, which is small on average but may vary as a function of source type, redshift, etc.

These comparisons establish that the PAE and TF probe broadly similar likelihood surfaces, with some systematic differences. For well-detected sources the differences are small, however for low-SNR objects they lead to substantially broader PAE profile likelihoods that more faithfully reflect the lack of constraining power in the data. This explains the diagnostic power of the $\sigma_{z/(1+z)}^{\rm PAE}/\sigma_{z/(1+z)}^{\rm TF}>2$ from \S \ref{sec:compare_redshift_errors}: flagged sources are cases where TF profiling produces artificially peaked $p(z)$ estimates despite an uninformative likelihood surface. In this regime, informative priors are necessary to break degeneracies in the data, as detailed in \S \ref{sec:prior_impact}. 

%% file: sections/sbi.tex
\section{Amortized redshift posterior estimation with Simulation-Based Inference}
\label{sec:sbi}
While we design a PAE that is intended for high-performance, scalable forward modeling, it may be desirable to have an even faster alternative when analyzing massive datasets with $\mathcal{O}(10^{8}\text{--}10^{9})$ measured galaxies. Simulation-based inference (SBI) has found many applications in cosmological inference \citep{Alsing2018, Cranmer2020, lemos2024field} and photometric redshift estimation \citep{Hahn2022}, and conditional normalizing flows in particular offer a natural framework for learning the posterior $p(z\mid\mathbf{x})$ directly from simulated data \citep{Papamakarios2021}. Once trained, the SBI network produces posterior samples via two deterministic steps: (1)~a single forward pass through the encoder to obtain a conditioning context vector, and (2)~transforming draws from a standard normal base distribution through the learned flow conditioned on that context. Because both steps are fully parallelizable on a GPU with no iterative or sequential-chain overhead, the cost per galaxy is low. This represents a data-driven method with an implicit redshift prior defined by the training set.

While we use the same underlying SPHEREx dataset as the PAE, for SBI we train on the set of noisy, observed frame photometry (rather than higher-resolution rest frame SEDs). We augment our initial training sample with multiple noise realizations per galaxy. Our SBI network pairs a Transformer encoder with a one-dimensional conditional normalizing flow, following the spectral-token architecture proposed by \citet{Parker2024} in AstroCLIP. The encoder is a six-layer, eight-head Transformer ($d_\mathrm{model}=256$, pre-norm) that operates on SPHEREx's 102-band photometry. Each input galaxy is represented as a two-channel
sequence---\texttt{arcsinh}-compressed flux and the associated photometric
uncertainty---which is sliced into overlapping tokens of width 10 with stride 5. We use \texttt{arcsinh} compression to account for the high dynamic range of the spectrum. A learnable \texttt{[CLS]} summary token is prepended; after the Transformer layers and a final layer norm, its 256-dimensional output serves as the conditioning context. Ultimately, this overlapping patch scheme allows the transformer to focus on localized spectral features while preserving spectral context at token boundaries; the deeper, wider Transformer then attends to the variety of spectral features in order to extract a
rich low-dimensional summary in the separate \texttt{[CLS]} vector. This vector is then passed to an eight-step masked affine
autoregressive flow \citep[MAF;][]{Papamakarios2017} with hidden dimension 128 that models the conditional density $p(z\mid\mathbf{x})$. The full model contains ${\sim}\,6.1\times10^{6}$ trainable parameters.

We train our model by minimizing the negative log-likelihood of the true redshift under the flow. Optimization uses Adam \citep{Kingma2015} ($\mathrm{lr}=10^{-3}$, $\lambda=10^{-5}$) with early stopping (patience of 10 epochs) on the validation NLL.  We consider two brightness-aware sampling strategies: \emph{flat} sampling, which reweights mini-batches so that each brightness quintile is represented equally, and \emph{importance} sampling, which applies a quadratic upweighting of the brighter bins ($w\propto b^{2}$, where $b$ is the brightness-bin index) to focus capacity on the high SNR regime. Both variants converge within 15--20 epochs on a single NVIDIA A100 GPU. In general, we find improved results across our dataset using importance sampling, and so we present the results using that training configuration.

A key advantage of SBI for this application is rapid posterior inference. On a single A100, the trained SBI network draws 1024 posterior samples per galaxy at a throughput of ${\sim}\,9\times10^{6}$\,galaxies\,node-hr$^{-1}$. This is nearly a factor of $200\times$ faster than our PAE for the same number of inference samples ($\sim 5\times 10^4$\,galaxies\,node-hr$^{-1}$).

\begin{figure*}
    \centering
    \includegraphics[width=0.4\linewidth]{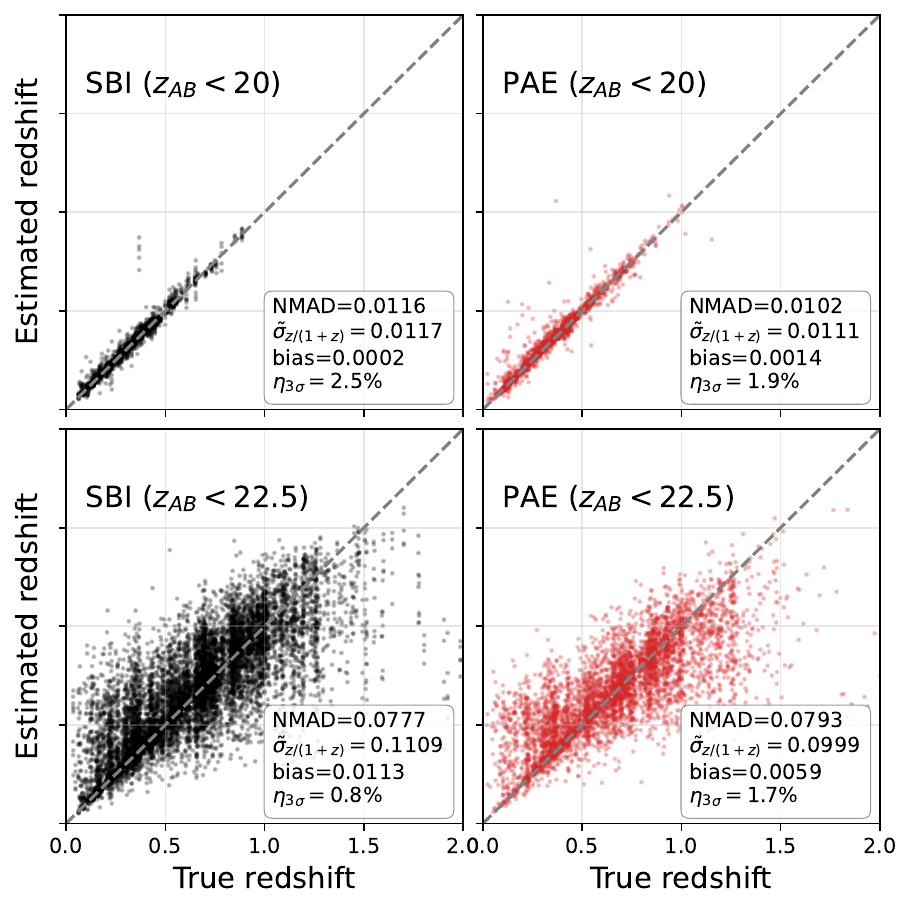}\includegraphics[width=0.235\linewidth]{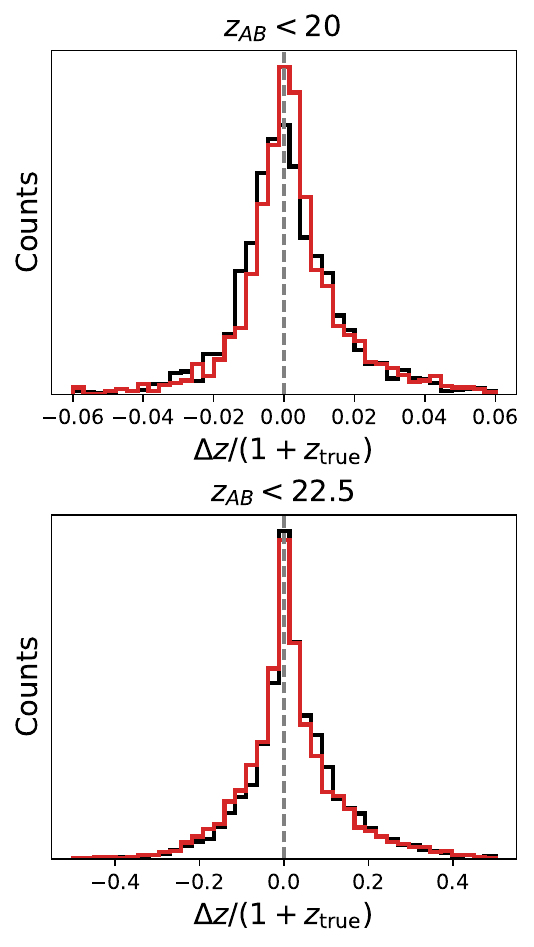}\includegraphics[width=0.3\linewidth]{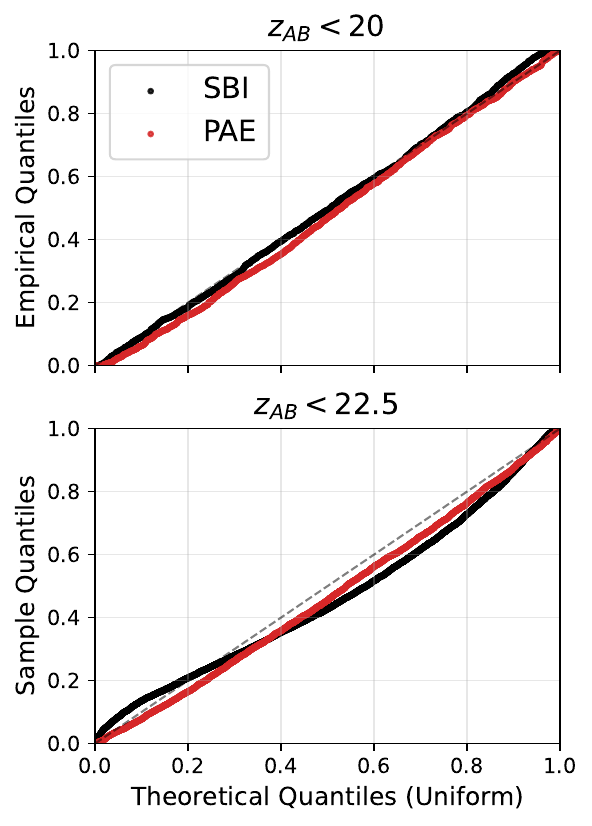}
    
    \caption{Comparison of simulation-based inference (SBI, black) and PAE (red) redshift performance, for bright ($z_{AB}<20$, top row) and full ($z_{AB}<22.5$, bottom row) samples. The left-hand plots show the recovered redshifts, the middle plots show the fractional redshift errors, and the right hand plots show the empirical coverage.}
    \label{fig:pae_sbi}
\end{figure*}

In Figure~\ref{fig:pae_sbi} we show the redshift recovery results from both methods, for two $z$-band limited samples.  For bright sources ($z_\mathrm{AB}<20$), SBI achieves $\mathrm{NMAD}=0.0116$ with a $3\sigma$ outlier fraction $\eta_{3\sigma}=2.5\%$ and negligible bias (0.0002), closely matching the PAE ($\mathrm{NMAD}=0.0102$, $\eta_{3\sigma}=1.9\%$).  For the fainter sample, $z_\mathrm{AB}<22.5$, SBI and the PAE are again comparable, with SBI obtaining $\mathrm{NMAD}=0.0777$ ($\eta_{3\sigma}=0.8\%$) versus $\mathrm{NMAD}=0.0793$ ($\eta_{3\sigma}=1.7\%$) for the PAE. Both methods recover redshift posteriors with nearly ideal empirical coverage for the $z_{\rm AB}<20$ cut, while for $z_{\rm AB}<22.5$ the coverage quality degrades mildly. Despite these relatively minor differences in performance between the PAE and SBI results, we conclude that both methods provide accurate and well-calibrated redshift estimates for a range of sources. 

%% file: sections/conclusion.tex
\section{Conclusion}
\label{sec:conclusion}

In this work, we have explored the use of probabilistic autoencoders as a basis for accurate and interpretable redshift estimation. In a controlled setting, we find that the PAE is competitive with template fitting, in some cases providing higher-precision estimates but, more importantly, yielding less biased, better-calibrated redshift posteriors. We have made our implementation, \texttt{PAESpec}, publicly available on Github\footnote{\url{https://github.com/RichardFeder/PAESpec}}, with instructions on how to train and adapt \texttt{PAESpec} to new datasets. 

The Bayesian forward modeling framework of \texttt{PAESpec} allows one to explicitly define priors that can be tailored to downstream science requirements. Our exploration of priors, both from PAE latent parameters and the redshift distribution reveal degeneracies that are not apparent in our TF baseline, particularly for low-SNR sources. The NF prior over latent parameters helps reduce the outlier fraction considerably by restricting the model space to physically plausible SEDs, while an informative redshift prior is necessary to break degeneracies for the faintest sources. By directly computing PAE profile likelihoods and comparing with TF, we find that the two methods probe likelihood surfaces with systematic differences arising from the discrete vs. continuous model structure. These profile likelihood comparisons provide an explanation for the diagnostic power of comparing PAE and TF uncertainties: the flagged sources identified in \S \ref{sec:pae_tf_errors} are those where TF profiling yields artificially confident redshift estimates despite an uninformative likelihood surface. Whether broader PAE posteriors also reflect volume contributions from marginalizing over the continuous latent space remains an open question that requires systematic comparison of PAE profiled and marginalized $p(z)$ across a larger sample.

\texttt{PAESpec}'s explicit forward modeling structure opens up several extensions beyond the scope of this work, a few of which we highlight here. Most directly, \texttt{PAESpec} enables generic modeling to SPHEREx's native flux measurements, bypassing the flux homogenization onto fiducial filters required by pre-computed template grids \citep{crill25} and improving redshift accuracy for sources  with detectable emission lines \citep{feder23}. Emission lines are straightforward to incorporate within our framework, either implicitly through the training SEDs or explicitly via a parametric model such as a sum of Gaussians. Additional photometry from external surveys can be folded in naturally, with relative zero-point offsets marginalized over as in \cite{stein_pae}. Looking further ahead, \texttt{PAESpec} could be used to simultaneously constrain the SEDs of spatially blended sources, though in that limit the flux covariance as a function of angular separation and wavelength becomes an important modeling ingredient.

A key assumption underlying the results of this work is that the training and test data are drawn from the same underlying SED library, providing a controlled setting in which differences in performance can be cleanly attributed to modeling choices rather than distribution shift. Sources that are poorly represented (e.g., high redshift sources, AGN, extreme emission line galaxies, etc.) may fall outside of the learned SED manifold, producing poor reconstructions or biased redshifts. For SPHEREx and other cosmology surveys with dedicated deep fields, high-quality multi-band photometry and existing spectroscopic redshifts provide a foundation for building representative training sets and calibrating the model against real data.  


Obtaining full Bayesian posteriors for sources is a more computationally intensive task than what is done in brute-force TF methods. Nonetheless, we have made significant progress in our implementation to deploy \texttt{PAESpec} at scale. For runs on 102-band photometry in which we deploy four chains per source with burn-in, hyperparameter tuning and 2000 steps per chain for posterior sampling, we achieve a processing rate of $\sim 4\times 10^4$ galaxies per node hour. We anticipate further improvements in processing throughput at scale from a combination of improved burn-in, fixed tuning based on empirical scaling relations, and shorter inference runs that leverage information on the autocorrelation length of high-/low-SNR sources.

To compare our results with a more ``data-driven" method, we considered a faster, SBI alternative to redshift estimation. When trained on noisy observed photometry, our SBI method yields competitive, well-calibrated photo-zs and offers a nearly 200$\times$ speed-up compared to our PAE. While these results are encouraging as an initial demonstration, the success of SBI methods for redshift estimation on real data will rely on high-fidelity training data with survey realism \citep{simbig}, diagnostics for model misspecification \citep{robust_sbi_lemos}, and thorough validation tests \citep{natali_sbi}. SBI may be most effective when developed and tested with the products of source injection pipelines, which are becoming a central element for the science data modeling and analysis toolkit of many surveys, including DES \citep{balrog_desy6}, DESI \citep{obiwan}, SPHEREx \citep{bock25},  and \emph{Rubin} \citep{rubin_ivezic}. SBI may be used as a forward scout for more intensive survey analyses, offering rapid tests of consistency between the recovered properties of real vs. injected sources that can be used to potentially diagnose and rectify instances of data/model mismatch relevant to selection function modeling relying on injections. 

The three estimation techniques we compare in this work capture some of the diversity within a much broader landscape of photo-z methods. In this context, the systematic evaluation of photo-z methods with rigorous and well-motivated metrics will play a central role in harnessing the full power of modern galaxy surveys that probe rich cosmological information through 3D clustering, gravitational lensing, and an ever-growing suite of cross-correlation observables.

%% file: sections/acknowledgements.tex
\acknowledgments

The authors thank Kangning Diao, Joshua Speagle and John Franklin Crenshaw for helpful discussions, Jakob Robnik and Minas Karaminas for help integrating MCLMC and \texttt{pocoMC}, and Zhaoyu Huai, Yun-Ting Cheng, Jean Choppin de Janvry, Yongjung Kim, Bomee Lee, and Dan Masters for comments on the manuscript. This work is supported by NASA TCAN (grant number 80NSSC24K0101). L.P. is supported by the NSF GRFP.

\software{\texttt{matplotlib} \citep{matplotlib}, \texttt{numpy} \citep{numpy}, \texttt{JAX} \citep{jax}, \texttt{BlackJAX} \citep{cabezas2024blackjax}, \texttt{FlowJAX} \citep{ward2023flowjax}, \texttt{equinox} \citep{equinox}}

%% file: sections/convergence_app.tex
\section{Convergence and effective sample size}
\label{sec:convergence}
\begin{figure}[h]
    \centering
    \includegraphics[width=0.48\linewidth]{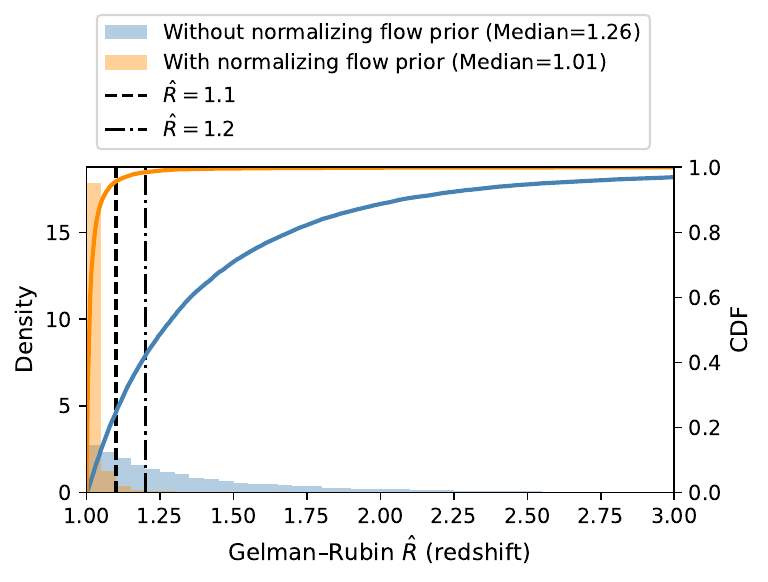}\includegraphics[width=0.49\linewidth]{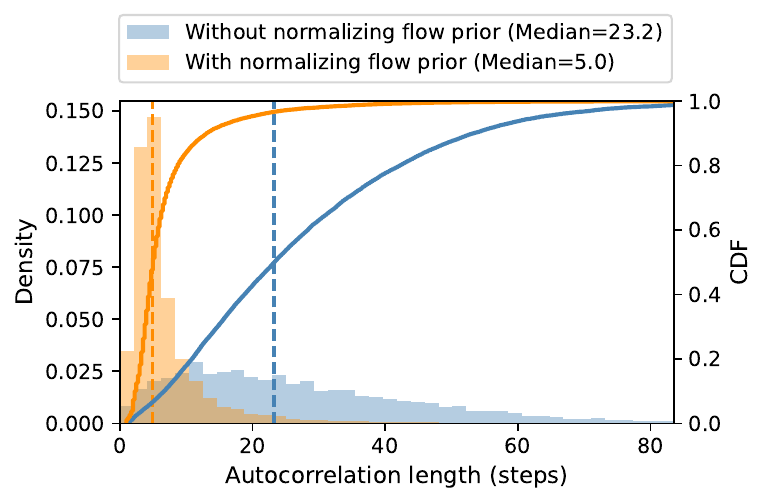}
    \caption{Distribution of the Gelman-Rubin statistic (left) and autocorrelation length (right) across galaxies for redshift using MCLMC. The twin axes show the cumulative distribution function (CDF) for both quantities.}
    \label{fig:rhat_acl_plots}
\end{figure}

In Figure \ref{fig:rhat_acl_plots} we show the distribution across sources for the Gelman-Rubin statistic $\hat{R}$, which compares the variance within and across chains, and the mean chain autocorrelation length (ACL). We estimate the ACL $\tau$ using the Sokal automatic windowing estimator with $c=5$ applied to the redshift dimension of each chain. By comparing runs on the same dataset, we find that the PAE normalizing flow prior has a significant impact on the convergence and mixing of our chains. In particular, the median (and 95th percentile) are 5.0 (13.0) with NF prior applied, compared to 23.0 (53.2) without. Likewise, 95.6 (98.5) per cent of sources have $\hat{R}<1.1$ (1.2) with the NF prior, compared to only 24.7 (42.1) per cent without the NF prior. We find that the redshift prior from \S \ref{sec:prior_impact} improves $\hat{R}$ slightly but does not change the qualitative results regarding the NF prior. These convergence non-idealities may impact the higher redshift outlier fractions observed in \S \ref{sec:prior_impact}, and further motivate using the NF prior in posterior redshift estimation. 

We conclude that our fiducial configuration has excellent convergence statistics -- for four chains with 1000 inference samples each, the median effective sample size (ESS) is $\sim 800$, i.e., Monte Carlo errors should be negligible for nearly all sources. We do find a trend between ACL and object SNR, implying a slightly lower ESS for the higher-precision sample. Future production runs may utilize this information, with an optimal performance strategy involving batching sources by SNR and varying the number of inference steps across batches to ensure sufficient sampling quality.

\section{Redshift performance for varying latent dimension}
\label{sec:vary_nlatent}
We investigate the sensitivity of our redshift recovery result to the dimensionality of the PAE latent space, denoted $n_{\rm latent}$. While increasing $n_{\rm latent}$ reduces SED reconstruction MSE (see Fig. \ref{fig:mse_recon_alltrain}), its impact on redshift estimation is less obvious. While additional latent dimensions may accommodate a richer SED manifold, they also introduce larger posterior volumes and potentially more sampling complexity. 

For each configuration ($n_{\rm latent}\in \lbrace 3, 5, 8, 10 \rbrace$), we run the PAE on the same set of 20,000 galaxies as in \S \ref{sec:results}, using both the NF and redshift priors established in the main text. Table \ref{tab:nlatent_redshift_performance} summarizes the key redshift performance metrics across runs, for one high-precision ($\sigma_{z/1+z}<0.01$) and one broader ($\sigma_{z/1+z}<0.2$) redshift uncertainty selection. 

The most significant performance differences arise in the high-precision selection -- we find that $n_{\rm latent}=3$ model yields a substantially elevated outlier fraction ($\eta_{3\sigma}=6.1\%$) and recovers the fewest high-precision sources. We note as well that the $n_{\rm latent}=3$ exhibits poorer chain convergence ($\hat{R}<1.1$ for 90.1\% of sources, compared to $\sim 96-98\%$ for larger $n_{\rm latent}$), suggesting that the lower capacity SED manifold has more complex posterior structure that is harder to sample, which may be coupled with the higher reconstruction MSE. As we increase $n_{\rm latent}$, we see improvement in nearly all metrics, and for $n_{\rm latent}=10$ we find the lowest outlier fraction $(\eta_{3\sigma}=1.6\%)$ and bias ($\langle \Delta z/(1+z) \rangle = 0.0004$). For the broader $\sigma_{z/1+z}<0.2$ selection, differences across models are smaller, though the outlier fraction decreases with $n_{\rm latent}$ and NMAD shows a modest improvement. From these results, we conclude that $n_{\rm latent}=10$ provides the best combination of redshift performance and posterior convergence, and we adopt it as our fiducial choice in \S \ref{sec:results} onward. 

It is possible that going to even higher $n_{\rm latent}$ would improve our results further. However, we caution that interpretation of the ``ideal" $n_{\rm latent}$ may change between simulated and real data, noting that the highly multimodal structure of the PAE latent space in Fig. \ref{fig:latents_dist} is partly an artifact of the discrete template set used to generate the SEDs. Real galaxy properties (star formation history, dust attenuation, metallicity, etc.) vary continuously, and so we posit that the true SED manifold is likely smoother and lower-dimensional than implied by our simulations. In other words, using higher $n_{\rm latent}$ may improve performance on our simulated sample by accommodating this discretized structure, while the optimal latent dimension for real data could be lower. We therefore view the latent dimension as a hyperparameter that can (and should) be validated on real data.


\begin{table*}

\hspace{-1cm}
\begin{tabular}{c c cccc cccc}
\hline\hline
& & \multicolumn{4}{c}{$\hat{\sigma}_{z/(1+z)} < 0.01$} & \multicolumn{4}{c}{$\hat{\sigma}_{z/(1+z)} < 0.2$} \\
\hline
PAE $n_\mathrm{latent}$ & $f(\hat{R}<1.1)$ [\%] & $N_\mathrm{src}$ & Bias & NMAD & $\eta_{3\sigma}$ [\%] & $N_\mathrm{src}$ & Bias & NMAD & $\eta_{3\sigma}$ [\%] \\
\hline
3  & 90.1 & 522 & $-0.0006$ & $0.0075$ & $6.1$ & 19521 & $0.0055$ & $0.0814$ & $2.1$ \\
5  & 96.3 & 720 & $0.0008$  & $0.0057$ & $2.4$ & 19825 & $0.0061$ & $0.0779$ & $1.6$ \\
8  & 96.1 & 645 & $0.0005$  & $0.0061$ & $2.2$ & 19737 & $0.0066$ & $0.0787$ & $1.5$ \\
10 & 97.7 & 707 & $0.0004$  & $0.0059$ & $1.6$ & 19787 & $0.0061$ & $0.0772$ & $1.4$ \\
\hline
\end{tabular}
\caption{PAE redshift recovery performance metrics for models with varying latent dimension. We compare high-precision (left) and low-precision (right) subsets of the PAE results. The quantity $f(\hat{R}<1.1)$ denotes the fraction of sources with Gelman-Rubin statistic $\hat{R}<1.1$.}
\label{tab:nlatent_redshift_performance}
\end{table*}